\documentclass[twocolumn,journal,10pt]{IEEEtran}
\usepackage[draft]{hyperref}

\usepackage[utf8]{inputenc}
\usepackage{glossaries}
\usepackage[symbols,nogroupskip]{glossaries-extra}
\usepackage{bm}
\usepackage{amssymb}
\usepackage{amsmath}
\usepackage{amsfonts}
\usepackage{mathrsfs}
\usepackage{bbm}
\usepackage{scalerel} 
\usepackage{graphicx}
\usepackage{subfigure}
\usepackage{cite}
\usepackage{enumerate}
\usepackage{url}
\usepackage{epstopdf}
\usepackage{algorithm}
\usepackage[noend]{algpseudocode}
\usepackage{algorithmicx,algorithm}
\usepackage{float}
\usepackage{subfigure}
\usepackage{amsthm}
\usepackage{color}
\usepackage{enumitem}
\usepackage{clipboard}
\usepackage[normalem]{ulem} 
\usepackage{framed}
\usepackage{array} 
\usepackage{hhline}
\usepackage[table]{xcolor}
\usepackage{balance}

\usepackage[colorinlistoftodos,prependcaption,textsize=scriptsize]{todonotes}
\usepackage{xargs} 
\usepackage{cooltooltips}
\usepackage{xcolor}  
\usepackage{comment}

\newcommand{\beq}{\begin{equation}}
\newcommand{\eeq}{\end{equation}}
\newcommand{\beqq}{\begin{equation*}}
\newcommand{\eeqq}{\end{equation*}}
\newcommand{\mbeqa}{\begin{align}}
\newcommand{\meeqa}{\end{align}}

\makeatletter
\def\@eqnnum{{\normalfont \normalcolor \theequation}}
\makeatother

\makeatletter
\newcommand{\manuallabel}[2]{\def\@currentlabel{#2}\label{#1}}
\makeatother

\makeatletter
\newcommand{\customlabel}[2]{%
   \protected@write \@auxout {}{\string \newlabel {#1}{{#2}{\thepage}{#2}{#1}{}} }%
   \hypertarget{#1}{}
}
\makeatother

\newtheorem{proposition}{\bf Proposition}

\newtheorem{lemma}{\bf Lemma}

\newcommand{\blue}{\textcolor[rgb]{0,0,0}}

\allowdisplaybreaks

\newcommand{\rev}{\textcolor[rgb]{0,0,0}}
\newcommand{\revs}{\textcolor[rgb]{0,0,0}} 

\newcommand{\tran}{\top}

\newcommand{\Diag}{\bm{\mathrm{Diag}}}

\newcommand{\softmax}{\mathtt{Softmax}\!}

\renewcommand{\Pr}{\operatorname{Pr}}

\newcommand{\name}{X$^2$Track}

\newcommand{\nname}{X$^2$Track}

\newcommand{\oPr}{\varGamma}  

\newcommand{\sB}{\mathrm{B}}
\newcommand{\sDT}{\scriptscriptstyle\mathrm{DT}}

\newcommand{\sSC}{\scriptscriptstyle\mathrm{SC}}
\newcommand{\sRIS}{\scriptscriptstyle\mathrm{RIS}}

\newcommand{\sCSI}{\scriptscriptstyle\mathrm{CSI}}
\newcommand{\sUL}{\scriptscriptstyle\mathrm{UL}}
\newcommand{\sIn}{\mathrm{in}}
\newcommand{\sOut}{\mathrm{out}}

\newcommand{\sTM}{\scriptscriptstyle\mathrm{TM}} 
\newcommand{\sTx}{\scriptscriptstyle\mathrm{Tx}} 
\newcommand{\sRx}{\scriptscriptstyle\mathrm{Rx}}

\newcommand{\sQ}{\scriptscriptstyle\mathrm{Q}} 
\newcommand{\sK}{\scriptscriptstyle\mathrm{K}}

\newcommand{\sV}{\scriptscriptstyle\mathrm{V}}

\newcommand{\sCE}{\mathrm{CE}}

\newcommand{\sTP}{\scriptscriptstyle\mathrm{TP}}

\newcommand{\stgt}{\mathrm{tgt}}
\newcommand{\ssrc}{\mathrm{src}}

\newcommand{\scen}{\mathrm{cen}}
\newcommand{\spre}{\mathrm{pre}}
\newcommand{\serD}{\scriptscriptstyle\mathrm{2D}}

\newcommand{\seql}{\left(}
\newcommand{\seqr}{\right)}

\newcommand{\attn}{\mathtt{Attn}}

\newcommand{\relu}{\mathtt{ReLU}}

\DeclareRobustCommand
\cpdots{\mathinner{\mkern-2mu ...\mkern-2mu}}

\begin{document}

\title{\linespread{1.2}\huge{Cross-domain Learning Framework for \\Tracking Users in RIS-aided Multi-band ISAC Systems \\ with Sparse Labeled Data}}

\author{\linespread{1.25}
\IEEEauthorblockN{
\normalsize{Jingzhi~Hu},~\IEEEmembership{\normalsize Member,~IEEE},
\normalsize{Dusit Niyato},~\IEEEmembership{\normalsize Fellow,~IEEE},
and~\normalsize{Jun~Luo},~\IEEEmembership{\normalsize Fellow,~IEEE}
}
\thanks{© 2024 IEEE.  Personal use of this material is permitted.  Permission from IEEE must be obtained for all other uses, in any current or future media, including reprinting/republishing this material for advertising or promotional purposes, creating new collective works, for resale or redistribution to servers or lists, or reuse of any copyrighted component of this work in other works.}
\thanks{This research is supported by National Research Foundation, Singapore and Infocomm Media Development Authority under its Future Communications Research \& Development Programme grant FCP-NTU-RG-2022-015. This work has been accepted for publication in the IEEE Journal on Selected Areas in Communications \emph{(Corresponding author: Jun Luo).}}
\thanks{
 J. Hu, D. Niyato, and J. Luo are with the College of Computing and Data Science, Nanyang Technological University, Singapore 639798~(email: jingzhi.hu@ntu.edu.sg, dniyato@ntu.edu.sg, junluo@ntu.edu.sg).
 }
}

\maketitle
\begin{abstract}
Integrated sensing and communications~(ISAC) is pivotal for 6G communications and is boosted by the rapid development of reconfigurable intelligent surfaces~(RISs).
Using the channel state information (CSI) across multiple frequency bands, RIS-aided multi-band ISAC systems can potentially track users’ positions with high precision.
Though tracking with CSI is desirable as no communication overheads are incurred, it faces challenges due to the multi-modalities of CSI samples, irregular and asynchronous data traffic, and sparse labeled data for learning the tracking function.
This paper proposes the \nname\ framework, where we model the tracking function by a hierarchical architecture, jointly utilizing multi-modal CSI indicators across multiple bands, and optimize it in a cross-domain manner, tackling the sparsity of labeled data for the target deployment environment (namely, target domain) by adapting the knowledge learned from another environment (namely, source domain).
Under \nname, we design an efficient deep learning algorithm to minimize tracking errors, based on transformer neural networks and adversarial learning techniques.
Simulation results verify that \nname\ achieves decimeter-level axial tracking errors even under scarce UL data traffic and strong interference conditions and can adapt to diverse deployment environments with fewer than $5\%$ training data, or equivalently, $5$~minutes of UE tracks, being labeled.

\end{abstract}

\section[Introduction]{Introduction}
\label{sec: intro}

Recently, research on 6G communications has been launched to achieve smart connectivity for everything~\cite{Liu2023WCM_Integrated}.
While various technologies are investigated to achieve this goal, a consensus has been developed that \emph{sensing} will evolve from an additional function to an essential necessity~\cite{Zhu2023SCIS_Pushing}, which makes the technique of integrated sensing and communications~(ISAC) of critical importance~\cite{Liu2022COMST_ASurvey}.
In addition to the intuitive idea of sharing spectrum resources and infrastructure hardware, ISAC also allows for sharing information in communications to support sensing functions~\cite{Liu2023WCM_Integrated}.
In this regard, the rapidly growing interest in ISAC can also be attributed to the emergence of reconfigurable intelligent surfaces~(RISs)~\cite{Liu2023WCM_Integrated, chepuri2022integrated}.
RISs enhance wireless communications by creating numerous controllable reflection paths and provide enriched channel information that can effectively bolster sensing functions.
The practicability of RISs stem from their cost- and power-efficient implementation of numerous controllable metamaterial antenna elements~(or, \emph{meta-elements} in short), yielding high controllability over reflected wireless signals~\cite{Pan2021COMMAG_Reconfigurable}. 

A remarkable advantage of RISs is their adaptive design approaches that can cater to diverse frequency bands, ranging from several GHz~\cite{tapio2021survey} to hundreds GHz~\cite{Tekbiyik2022IEEEAcc_Reconfigurable}, even to visible light~\cite{Cao2020VTC_Reflecting}. 
This adaptability makes RISs especially suitable for 6G, where jointly utilizing multiple frequency bands to deliver ISAC services is essential~\cite{Wang2020VTM_6G}.
Existing work has  explored the design and applications of RIS-aided single-band communication and sensing systems in sub-6~\!GHz~\cite{li2019machine}, mmWave~\cite{Tang2022TCom_Path}, and THz~\cite{Wan2021TCOM_Terahertz}. 
Several studies have investigated dual-band RIS systems, aiming at enhancing communication rates\cite{Jiang2022JCN_AJoint, cho2022towards,Lin2022JPhyD_ADual} without discussing their sensing potentials.
Moreover, few of them consider RIS-aided ISAC systems working in multiple bands~\cite{Hu2023ICC_Multi,Hu2023JSAC_HoloFed}.
Despite the limited existing work, RIS-aided multi-band~(MB) ISAC systems are of high research value due to the complementary nature of different bands, which can endow communications with high-resolution and wide-coverage sensing capability~\cite{Liu2023WCM_Integrated}.

To fill the research gap, we take the first step to investigate the RIS-aided MB ISAC systems.
We focus on unleashing their potential for user equipment~(UE) tracking, which is crucial for supporting various location-based services and applications~\cite{Chen2022IoTJ_Reconfigurable, Strinati2021_6GSUMMIT_Wireless}.
Different from conventional studies on RIS-aided ISAC systems, where the RISs generate separate beams~\cite{Zhang2022Holographic, chepuri2022integrated,Xing2023TCOM_Joint}, 
	alternate between sensing and communication modes~\cite{zhang2021metalocalization,Nguyen2021IEEEAcc_Wireless,Jiang2022arxiv_Optimization, Keykhosravi2023VTM_Leveraging,He2020WCNC_Adaptive}, 
	or require additional sensor devices\cite{Tardif2023WCM_SelfAdaptive, Hu2022TWC_IRS}, 
	we directly utilize the channel state information (CSI) readily available in communications.
By this means, we ensure uncompromised quality of services for communication.
\rev{More specifically, we focus on the CSI indicators of the channel gain between each Tx and Rx antenna pair over multiple subcarriers, which represent the spectrum of the channel impulse response~(CIR) of the time domain.}
\rev{We note that owing to their numerous meta-elements, RISs essentially have a large aperture size~\cite{ElMossallamy20TCCN_RIS} and thus enhance the sensing range and resolution by enabling the MB CSI to be measured in a large space region with fine granularity.}

The approach of directly using CSI for sensing is in line with the concept of \emph{perceptive mobile networks}~(PMNs), which has become increasingly popular in 6G research~\cite{Zhang2021VTM_Perceptive, Xie2023WCM_Collaborative, Zhang2022COMST_Enabling} and Wi-Fi based ISAC systems~\cite{ChenCOMM23_ISACoT, Hu2023Mobicom_Muse, Li2024Mobisys_UWB}.
However, only a few studies~\cite{Wang2022JSAC_Location, Shao2022Arxiv_Joint} have explored the role of RISs in PMNs.
In~\cite{Shao2022Arxiv_Joint}, the authors considered a specific scenario involving a full-duplex base station~(BS) and multiple UEs equipped with on-device RISs. 
A more general analysis is provided in~\cite{Wang2022JSAC_Location}; however, it is derived from a theoretical perspective under simplified assumptions, \blue{e.g., no strong reflection or scattering multipath interference}, leaving the challenges of practical data traffic and wireless channel conditions unaddressed, especially in MB ISAC systems.

\rev{In distinct contrast to prior work, which focus on single-band RISs, require extra sensing signals, or apply to specific scenarios with low interference, we consider an RIS-aided MB ISAC system in practical wireless environments with high interference, leveraging readily available CSI in uplink~(UL) traffic for sensing. 
In the considered system, we enable the BS to learn a UE tracking function while tackling the challenges of highly complex MB CSI indicators, practical UL traffic conditions, and limited knowledge of deployment environments.}
In particular, the three most prominent challenges can be elaborated on as follows:
\begin{itemize}[leftmargin=*]
\item \textbf{Multi-modality of CSI Indicators}: 
	In RIS-aided MB ISAC systems, CSI indicators are from multiple bands and comprise the channel gains for direct transmission paths between Tx and Rx antennas and for reflection paths via the RISs\cite{Cao2021COMMAG_AI, Swindlehurst2022ProcIEEE_Channel}, which are of distinct physical meanings and data modalities and challenging to be jointly utilized. 
\item \textbf{Irregularity and Asynchrony of Data Traffic}:
	Due to generally unsaturated UL traffic and independent scheduling of each band, the samples of CSI indicators are irregularly spaced in time, and the CSI sequences of multiple bands are asynchronous, which are challenging to handle~\cite{Narayan2021ICLR_Multi,Zhang2022ICLR_Graph}.
\item \textbf{Sparsity of Labeled Data for Learning}: 
Owing to the diversity and complexity of deployment environments, collecting sufficient CSI indicator data labeled with ground-truth UE tracks to learn the tracking function incurs a prohibitive cost. In general, only sparse labeled data can be obtained, making it challenging to learn an accurate tracking function with enough generalizability.
\end{itemize}

To handle the above challenges, we propose a novel framework named \nname. 
The core of \nname\ is to model the tracking function in a hierarchical architecture, where the CSI indicators \emph{across} multiple modalities are first encoded and aggregated at per-frame level and then go through mutual complementation at the sequence level.
Moreover, \nname\ optimizes the tracking function for tracking error minimization in a \emph{cross}-domain manner, tackling the sparsity of labeled data for the target deployment environment~(namely, target domain) by utilizing available data in a different known environment~(namely, source domain).
Under \nname, we implement an efficient deep learning algorithm, where we design a transformer-based deep neural network~(DNN) for the tracking function and employ adversarial learning techniques to solve the cross-domain tracking error minimization.
Our main contributions are summarized below.
\begin{itemize}[leftmargin=*]
\item We are the first to investigate the potential of RIS-aided MB ISAC systems for tracking UEs solely with the CSI indicators available in communications. We establish the channel model and propose a protocol to specify how CSI indicators are obtained and used for track estimation. 
\item We propose the \nname\ framework, handling the challenges including multi-modality of CSI indicators, irregularity and asynchrony of traffic, and sparsity of labeled data for learning, and design an efficient algorithm under \nname\ for tracking error minimization.
\item Simulation results verify that \nname\ achieves decimeter-level axial tracking errors given scarce UL traffic and strong interference and enables generalizability to diverse environments with less than $5\%$ CSI data being labeled, equivalently, $5$~minutes of UE tracks.
\end{itemize}

The rest of this paper is organized as follows:
In Sec.~\ref{sec: system model}, we establish the model of the RIS-aided MB ISAC system.
In Sec.~\ref{sec: prob formulation}, we formulate the tracking error minimization problem for UE tracking, analyze its challenges, and propose the \nname\ framework to handle the challenges.
In Sec.~\ref{sec: alg design}, we design an algorithm under the \nname\ framework to solve the tracking error minimization problem.
Simulation results are provided in Sec.~\ref{sec: evaluation}, and a conclusion is drawn in Sec.~\ref{sec: conclu}.

\emph{Notations}: 
$\mathbb R^{M\times N}$, and $\mathbb C^{M\times N}$ denote the sets of real and complex $M\times N$ matrices, respectively. 
$\|\cdot \|_1$ and $\|\cdot\|_2$ are the $\ell^1$-norm and $\ell^2$-norm, respectively.
$\circ$ represents the function composition, $\odot$ denotes the Hadamard product, and $\oplus$ is the array concatenation along the last dimension.
\revs{$(\cdot)^\tran$ denotes the transpose operation.}
$\Diag(\bm x)$ returns the matrix whose main diagonal vector is $\bm x$, with other elements being zeros.
$\mathbb E_{\bm x\sim \varGamma}(\cdot)$ returns the expectation of the argument, given that variable $\bm x$ follows distribution~$\varGamma$.
\revs{$\sup_{\bm f\in\mathcal X} \epsilon(\bm f)$ returns the supremum of $\epsilon(\bm f)$ for all $\bm f$ in set $\mathcal X$.}
$\mathcal J =\seql\bm x_i|\mathcal H_i=1\seqr_i$ is the sequence of $\bm x_i$ for all feasible subscripts $i=1,2,\cpdots$, where $\bm x_i$ satisfies hypothesis $\mathcal H_i=1$.
For sequences $\mathcal J_1$ and $\mathcal J_2$, $\|\mathcal J_1 - \mathcal J_2\|_1$ denotes the sum of the $\ell^1$-norm values of their element-wise differences.
Given $\mathcal J$ being a sequence or a set, $|\mathcal J|$ represents the number of elements in $\mathcal J$.
$[\bm x]_{m}$, $[\bm X]_m$, and $[\bm X]_{m,n}$ denote the $m$-th element of vector $\bm x$, the $m$-th row vector of matrix $\bm X$, and the $(m,n)$-th element of matrix $\bm X$, respectively.
$\bm {\mathsf X}$ represents a multi-dimensional array with more than three dimensions, and $[\bm{\mathsf X}]_{m,n\cpdots}$ denotes the block element of $\bm {\mathsf X}$ with the indices of its first dimensions being $m,n,\cpdots$.
\revs{Finally, given $x\in\mathbb C$, its amplitude and angle are denoted by $|x|$ and $\angle x$, respectively.}

\section[System Model]{Model of RIS-Aided MB ISAC Systems}
\label{sec: system model}

\begin{figure}[t]  
\centering
\includegraphics[width=1\linewidth]{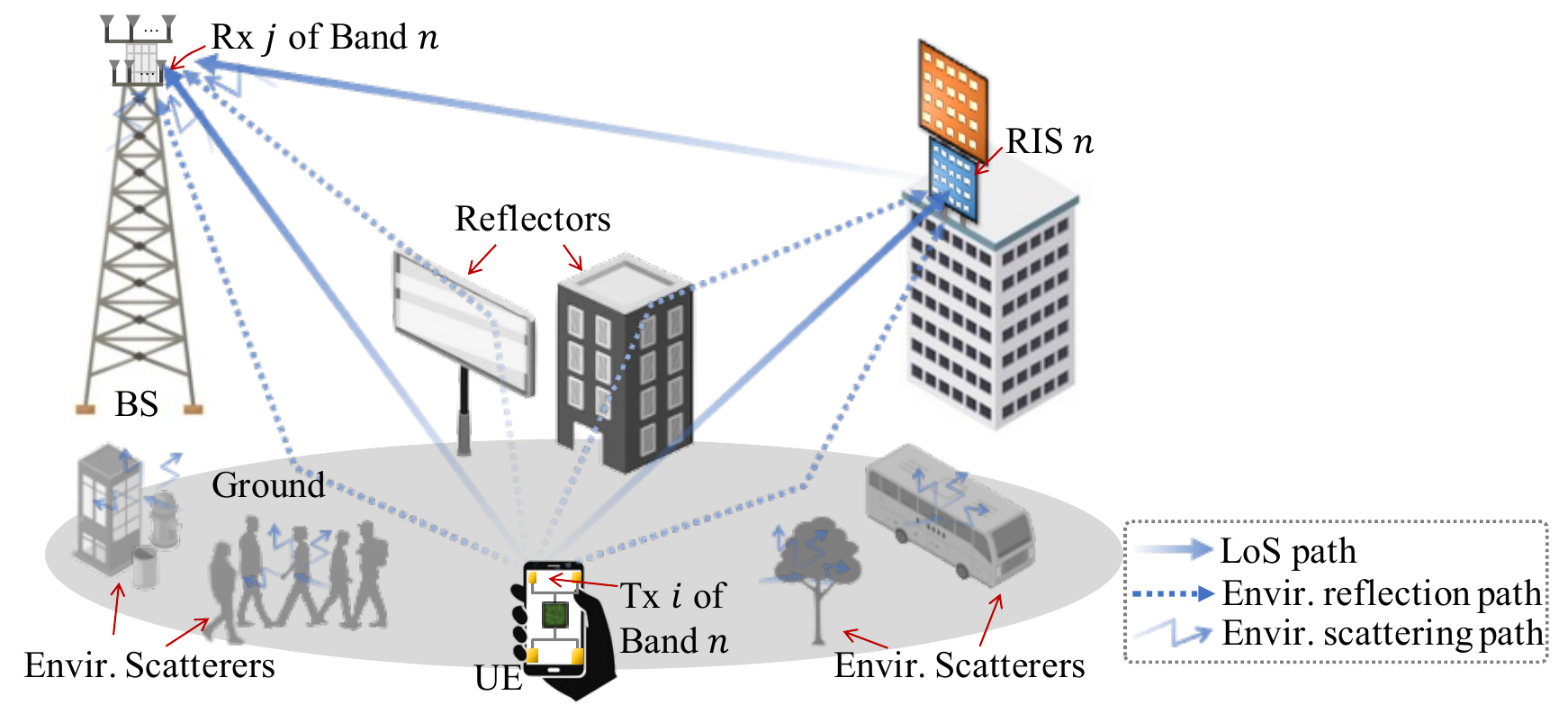}
  \vspace{-1em}
  \caption{An RIS-aided MB ISAC system and the wireless channel for the UL transmission from a UE to the BS in Band $n$.}
  \label{fig: sys channel model}
\vspace{-1em}
\end{figure}

As shown in Fig.~\ref{fig: sys channel model}, an RIS-aided MB ISAC system comprises multiple mobile UEs, a BS that provides UL communication services via $N_{\sB}$ frequency bands to the UEs within a 3D region of interest~(ROI)~$\mathcal A\subset \mathbb R^{3}$, and \rev{$N_{\sRIS}$ RISs~($N_{\sRIS}=N_{\sB}$)} each working in a different band and aiding the communications between the BS and UEs.
\revs{The number of sub-carriers of Band $n$ is denoted by $N_{\sSC,n}$.}
For the UL communications in Band $n$, the BS is equipped with $N_{\sRx,n}$ Rx antennas, and each UE is equipped with $N_{\sTx, n}$ Tx antennas.
Besides, the RIS for Band $n$, i.e., RIS $n$, is composed of $M_n$ densely arranged meta-elements, and the size of each meta-element is $\gamma\in[0,1]$ fraction of the center wavelength of Band $n$.
Each meta-element imposes a controllable reflection response to its reflected signals, and the reflection coefficients of all $M_n$ meta-elements constitute the \emph{configuration} of RIS~$n$.
By controlling the configuration of RIS~$n$, the BS enables the RIS $n$ to perform passive beamforming and focus UE's transmission signals towards the BS's Rx antennas.

During the UL transmission from an arbitrary UE to the BS, the BS collects multiple CSI indicators, such as the CSI indicators for the direct transmission~(DT) from Tx to Rx antennas and for the reflection paths via the RISs~\cite{Cao2021COMMAG_AI, Swindlehurst2022ProcIEEE_Channel}.
Using the CSI indicators collected in multiple bands, the BS provides tracking services for the UEs.
We assume that the tracking is performed periodically, where each period is referred to as a \emph{tracking period} and of duration $T_{\sTP}$.
The BS employs a \emph{tracking function} denoted as $\bm f$ to map the sequences of CSI indicators in each tracking period to the estimation of UEs' tracks.
As the tracking period is the same for each UE, we will focus on tracking an arbitrary UE throughout the remainder of this paper. 
The central position of the considered UE at time $t$ is represented by $\bm p(t)\in\mathcal A$.

In the following, we establish the channel model to facilitate the description of multi-modal CSI indicators and then explain how these CSI indicators are obtained, formatted, and utilized for UE track by proposing an RIS-aided MB ISAC protocol.

\subsection{Channel Model}
\label{ssec:channelmod}

As shown in Fig.~\ref{fig: sys channel model}, we assume that the system is targeted for deployment in a general urban area with abundant environmental scatterers (e.g., pedestrians, trees, and street accessories) and many reflectors (e.g., ground, billboards, building surfaces) for wireless signals, constituting an interference-rich environment for wireless transmissions.
Following the channel modeling approach of the 3GPP standard~\cite{3GPP_TR_38_901}, we model the channel between the UE and the BS in a hybrid manner, combining deterministic and statistical models for line-of-sight~(LoS) paths, environmental reflection paths, and environmental scattering paths.

Specifically, we focus on the CSI indicators that comprise the complex channel gains for the subcarriers across the $N_{\sB}$ bands.
For each Band $n$~($n\in\{1,\cpdots,\!N_{\sB}\}$), we denote the $i$-th Tx antenna of the UE as Tx $i$ and the $j$-th Rx antenna of the BS as Rx $j$~$(i\in\{1,\cpdots,\! N_{\sTx,n}\}, j\in\{1,\cpdots,\! N_{\sRx,n}\})$.
At time $t\in[0,T_{\sTP}]$ within a tracking period, the channel gain for signals of frequency $f$ can be modeled as~\cite{Jiang2022JCN_AJoint}:
\begin{equation}
\label{equ: overall channel}
h_{n,i,j}(t,f) = h_{n,i,j}^{\sDT}(t,f) + h^{\sRIS}_{n,i,j}(t,f).
\end{equation}
Here, $h_{n,i,j}^{\sDT}(t,f)\in\mathbb C$ is the gain from Tx $i$ to Rx $j$ excluding the gain due to the reflection via RIS $n$~(referred to as the DT channel gain), and $h^{\sRIS}_{n,i,j}(t,f)\in\mathbb C$ represents the aggregate gain for the reflection paths via the $M_n$ meta-elements of RIS~$n$~(referred to as the RIS channel gain).

\subsubsection{DT Channel Gain}
\label{s3ec: dt channel gain}

\rev{Rather than using a channel fading coefficient to represent the collective influence of multiple channel fading factors, we distinctly model each factor instead.}
Based on~\cite{goldsmith2005wireless}, $h_{n,i,j}^{\sDT}(t,f)$ comprises the gains for the LoS path, environmental reflection paths, and other environmental scattering paths, i.e.,
\begin{align}
\label{equ: channel gain direct}
h_{n,i,j}^{\sDT}(t,\! f) =  &I^{\sDT}_n \cdot g(\|\bm a^{\sTx}_{n, i}(t)-\bm a^{\sRx}_{n,j}\|_2, f)\! \\
&+\sum_{r=1}^{R}\varOmega^{\sDT}_{n,r} \cdot g( D^{\sDT}_{n,r} (\bm a^{\sTx}_{n, i}(t), \bm a^{\sRx}_{n,j}) , f) + H^{\sDT}_{n,i,j}. \nonumber
\end{align}
\blue{In~\eqref{equ: channel gain direct}}, $I^{\sDT}_n\!\in\!\{0,1\}$ is an indicator for the existence of the LoS path between the UE's Tx antennas and the BS's Rx antennas for Band $n$ and can be modeled by a random variable following the Bernoulli distribution with probability $\Pr^{\sDT}_n\in[0,1]$.
Besides, $g(d, f)$ is the free-space propagation gain in~\cite{goldsmith2005wireless} for signals of frequency $f$ traveling distance $d$.
Then, $\bm a_{n,j}^{\sRx}\in\mathbb R^3$ denotes the location of Rx $j$, and \rev{$\bm a_{n,i}^{\sTx}(t)\in\mathbb R^3$ is the position of Tx $i$ at time $t$, which depends on UE's position $\bm p(t)$}.

Moreover, in the summation term of~\eqref{equ: channel gain direct}, $R$ denotes the number of environmental reflection paths, and $\varOmega_{n,r}^{\sDT}\in\mathbb C$ represents the reflection coefficient for the $r$-th path and is modeled as a random complex variable, i.e., $\varOmega_{n,r}^{\sDT}\sim\mathcal{CN}(0,V_n^{\sDT})$, where $V_n^{\sDT}$ represents the variance.
Furthermore, 
$D_{n,r}^{\sDT}(\bm a^{\sTx}_{n, i}(t), \bm a^{\sRx}_{n,j})\in\mathbb R$ is the traveling distance from $\bm a^{\sTx}_{n, i}(t)$ to $\bm a^{\sRx}_{n,j}$ over the $r$-th reflection path.
Due to the complexity and dynamics of urban environments, $D_{n,r}^{\sDT}(\bm a^{\sTx}_{n, i}(t), \bm a^{\sRx}_{n,j})$ is modeled as a random variable, exhibiting correlation across different $i$, $j$, and $t$.
\rev{More specifically, as multiple reflections incur large attenuation, we model this traveling distance by the distance from the transmitter to the receiver via a single reflection point that is evenly sampled within~$\mathcal A$.}

Finally, $ H^{\sDT}_{n,i,j}$ in~\eqref{equ: channel gain direct} represents the aggregate gain of the environmental scattering paths and is modeled by a random complex value following complex Gaussian distribution $\mathcal{CN}(0,\varLambda^{\sDT}_n)$ with $\varLambda^{\sDT}_n$ being its variance.
We note that $I^{\sDT}_n$ and $H^{\sDT}_{n,i,j}$ are dependent on the position of the UE, i.e., $\bm p(t)$, and their relationship with $\bm p(t)$ can be modeled based on the 3GPP standard~\cite{3GPP_TR_38_901}, which are elaborated on in Sec.~\ref{s2ec: simul set}.

\subsubsection{RIS Channel Gain}
\label{s3ec: ris refl channel gain}

Based on~\cite{Jiang2022JCN_AJoint}, we model $h^{\sRIS}_{n,i,j}(t,f)$ in~\eqref{equ: overall channel}, i.e., the channel gain from Tx $i$ to Rx $j$ via RIS $n$, as:
\beq
\label{equ: gain ris reflection}
h^{\scriptscriptstyle\mathrm{RIS}}_{n,i,j}(t,f) = \bm g^{\sRx}_{n,j}(t,f)^\tran \Diag(\bm\phi_n(t,f))\bm g^{\sTx}_{n,i}(t,f),
\eeq
where $\bm g^{\sTx}_{n,i}(t,f), \bm g^{\sRx}_{n,j}(t,f)\in\mathbb C^{M_n}$ are the gain vectors from Tx $i$ to the $M_n$ meta-elements and from the $M_n$ meta-elements to Rx $j$, respectively, and $\bm \phi_n(t,f)\in\mathbb C^{M_n}$ denotes the reflection coefficients of the $M_n$ meta-elements, i.e., the configuration of RIS $n$.

Similar to the model of DT channel gain in~\eqref{equ: channel gain direct}, the gain from Tx $i$ to the $m$-th meta-element~($\forall m\in\{1,\cpdots,\! M_n\}$) of RIS $n$ can be modeled as
\begin{align}
\label{equ: gain tx2ris}
[\bm g^{\sTx}_{n,i}&(t,f)]_{m} = 
I^{\sRIS}_n\!\cdot\! g(\|\bm a^{\sTx}_{n, i}(t)\!-\! \bm u_{n,m}\|_2, f) \\
& +\sum_{r'=1}^{R'} \varOmega_{n,r'}^{\sRIS}
\cdot 
g(D_{n,r'}^{\sRIS}(\bm a_{n,i}^{\sTx}(t), \bm u_{n,m}),f) 
+
H_{n,i,m}^{\sRIS}. \nonumber
\end{align}
Here, $I_n^{\sRIS}\in\{0,1\}$ follows the Bernoulli distribution with probability $\Pr_n^{\sRIS}$, where $\Pr_n^{\sRIS}$ denotes the probability of the LoS path between the UE and RIS $n$,
and $ \bm u_{n,m}\in\mathbb R^{3}$ denotes the location of the $m$-th meta-element of RIS~$n$.
Moreover, similar to $R$, $\varOmega_{n,r}^{\sDT}$, $V^{\sDT}_n$, and $D_{n,r}^{\sDT}(\cdot)$  in~\eqref{equ: channel gain direct}, $R'$, $\varOmega_{n,r'}^{\sRIS}$, $V^{\sRIS}_n$, and $D_{n,r'}^{\sRIS}(\cdot)$ are defined for the environmental reflection paths between the UE and RIS $n$.
Symbol $H_{n,i,m}^{\sRIS}\!\sim\! \mathcal{CN}(0,\varLambda_n^{\sRIS})$ is the aggregate gain for the environmental scattering paths from Tx $i$ to meta-element $m$, with $\varLambda_n^{\sRIS}$ being its variance.

As for $\bm g^{\sRx}_{n,j}(t,f)$ in~\eqref{equ: gain ris reflection}, we focus on the practical scenario widely considered in existing RIS-aided communication systems, where the BS and the RISs are deployed in open areas well-above the ground and have \blue{LoS paths between them dominating other multi-path channels~\cite{Liu2021ICC_Reconfigurable, Elzanaty2023IoTM_Towards,Cao2021COMMAG_AI, Liu2023WCM_Integrated,Swindlehurst2022ProcIEEE_Channel}.}
In this scenario, $\bm g^{\sRx}_{n,j}(t,f)$ can be expressed as
\!\footnote{\blue{In~\eqref{equ: gain reflection RIS}, reflection and scattering paths are omitted for the conciseness of presentation because they are dominated by the LoS path in the considered scenario.
Incorporating the gains of reflection and scattering paths into~\eqref{equ: gain reflection RIS} will not compromise the effectiveness of the proposed framework and algorithm in Secs.~\ref{sec: prob formulation} and~\ref{sec: alg design}, as they are not dependent on~\eqref{equ: gain reflection RIS}.}}
\newglossaryentry{angle_nj}{name={\ensuremath{\tilde{\bm \theta}_{n,j}}}, sort={}, description={}}
\beq
\label{equ: gain reflection RIS}
\left[\bm g^{\sRx}_{n,j}(t,f)\right]_{m} \!=\! g(\|\bm u_{n,m} \!- \! \bm a^{\sRx}_{n,j}\|_2, f),~\forall m\! \in\!\{1,\cpdots,M_n\}.
\eeq

\subsection{RIS-aided MB ISAC Protocol}
\label{s2ec: protocol}

In the following, we propose the RIS-aided MB ISAC protocol.
The proposed protocol is based on the existing ones for RIS-aided communication systems in~\cite{Cao2021COMMAG_AI, Liu2021COMST_Reconfigurable} and the 5G new radio~(NR) UL communication standards of 3GPP~\cite{3GPP_TS_38_212, 3GPP_TS_38_214}.  
We note that our proposed protocol achieves ISAC since all the CSI used for the estimation of UE track are readily provided in UL communications, and there are no overhead transmissions dedicated to positioning or tracking~(such as beam scanning or generating special beam patterns).
Therefore, it can readily co-exist with other co-channel communication systems, which is an important prerequiste for deployment in practical crowded wireless environments~\cite{muzi, favor, art}. %
\rev{Moreover, as the protocol only utilizes the functionality of RISs in aiding CSI collection, no optimization of the RISs'  configurations is necessary, making the protocol fully compatible with arbitrary control methods of the RISs.}

Within each tracking period, the time of Band~$n$~($\forall n\in\{1,\cpdots,\!N_{\sB}\}$) is discretized into frames with duration $T_n$, according to the 3GPP standard~\cite{3GPP_TS_138_211}.
Thus, the number of frames in a tracking period is $F_n=T_{\sTP}/T_n$, which is indexed by $\tau$~($\tau\in \{1,\cpdots, F_n\}$).
As illustrated in Fig.~\ref{fig: protocol design}, each frame comprises the following four phases:

\begin{figure}[t]  
\centering
\includegraphics[width=1.0\linewidth]{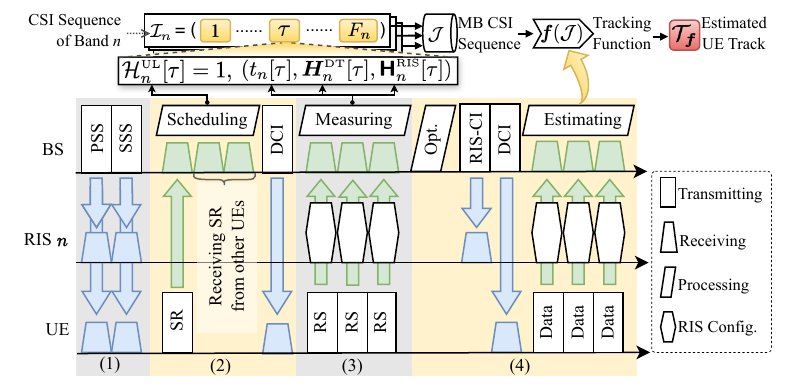}
  \vspace{-1em}
  \caption{\rev{Illustration of the RIS-aided MB ISAC protocol, including the components of MB CSI sequnces and the estimation of UE track.}}
  \label{fig: protocol design}
\end{figure}

(1)~\textbf{Synchronization}:
At the beginning of each frame, the BS, RIS $n$, and the UE synchronize their time and carrier frequency.
This is done by the UE and the controller of RIS $n$ receiving the downlink primary and secondary synchronization signals~(PSS and SSS) from the BS~\cite{Dahlman2020Book_5G}.

(2)~\textbf{Request and Scheduling}:
When the UE has UL data, it sends a scheduling request~(SR) to the BS via the UL control channel.
The BS schedules the requests from all the UEs and sends downlink control information~(DCI) to each UE via the downlink control channel, informing whether the user can transmit UL data in this frame.
When the UE is scheduled for UL transmission in Frame $\tau$ of Band $n$, it is represented by the hypothesis $\mathcal H_n^{\sUL} [\tau]=1$; otherwise, $\mathcal H_n^{\sUL} [\tau]=0$.

(3)~\textbf{Channel Sounding}:
Given $\mathcal H_n^{\sUL} [\tau]=1$, the UE transmits reference signals~(RS) in UL, during which RIS~$n$ controls its configurations to facilitate the channel sounding.
Based on~\cite{Cao2021COMMAG_AI, Swindlehurst2022ProcIEEE_Channel}, the BS estimates the DT channel gains among the Tx and Rx antenna pairs as well as the RIS channel gains via the $M_n$ meta-elements of RIS $n$, which is a pre-requisite for controlling the RIS's configuration to aid communications.
Therefore, in this sounding phase, the BS obtains two types of CSI indicators: DT-CSI, and RIS-CSI.
The DT-CSI is formatted as 2D complex matrix $\bm H^{\sDT}_n [\tau]\in \mathbb C^{Q_n\times N_{\sSC,n}}$, and the RIS-CSI is formatted as 3D complex array $\boldsymbol{\mathsf{H}}^{\sRIS}_n [\tau]\in \mathbb C^{Q_n\times N_{\sSC,n}\times M_n}$, where $Q_n = N_{\sTx,n}\cdot N_{\sRx,n}$ denotes the total number of channels formed by pairing the $N_{\sTx,n}$ Tx with the $N_{\sRx,n}$ Rx antennas.
Based on~\eqref{equ: channel gain direct} and~\eqref{equ: gain ris reflection}, denoting the indices of channels between Tx and Rx pairs and of sub-carriers by $q=(i-1)\cdot N_{\sRx,n} + j$~($q\in\{1,\cpdots,\! Q_n\}$) and $k$~($k\in\{1,\cpdots,\! N_{\sSC,n}\}$), respectively, the elements of $\bm H^{\sDT}_n [\tau]$ and $\boldsymbol{\mathsf{H}}^{\sRIS}_n [\tau]$ can be expressed  as
\begin{align}
\label{equ: csi format with noise}
& \big[\bm H^{\sDT}_n [\tau]\big]_{q,k} \!=\! h_{n,i,j}^{\sDT}(t_n[\tau], f_{n,k}) \! + \! e_n^{\sDT}, \\
&\big[\boldsymbol{\mathsf{H}}^{\sRIS}_n [\tau]\big]_{q,k} \!=\! \bm g^{\sRx}_{n,j}(t_n[\tau],f_{n,k})\odot \bm g^{\sTx}_{n,i}(t_n[\tau],f_{n,k}) \! +\!  \bm e_n^{\sRIS}, \nonumber 
\end{align}
where $t_n[\tau]$ represents the time of the channel sounding phase of Frame $\tau$ relative to the beginning of the tracking period, and $f_{n,k}$ is the frequency of Sub-carrier $k$ of Band $n$.
Besides, $e_n^{\sDT}\in\mathbb C$ and $\bm e_n^{\sRIS}\in\mathbb C^{M_n}$ denote the measurement noises due to thermal noises and measurement errors. 
\rev{We assume $e_n^{\sDT}\sim\mathcal{CN}(0, \sigma_n^{\sDT})$ and $\bm e_n^{\sRIS}\sim\mathcal{CN}(\bm 0, \sigma_n^{\sRIS}\bm I)$ with $\sigma_n^{\sDT}$ and $\sigma_n^{\sRIS}$ being the variances and $\bm I$ denoting the identity matrix.}

(4)~\textbf{Data Transmission and Tracking}: 
Based on the collected CSI indicators in the sounding phase, the BS optimizes~(opt.) its spectrum resource allocation and the configuration of RIS $n$.
It informs RIS $n$ and the UE by sending RIS Control Information~(RIS-CI) and DCI, respectively.
Then, aided by RIS $n$, the UE transmits its UL data to the BS.

At the end of a tracking period, the BS formats the collected CSI indicators as a sequence and estimates the track of UE based on it.
The collected CSI indicators include the DT-CSI and RIS-CSI for the $N_{\sB}$ bands.
In particular, in Frame $\tau$ of Band $n$, given $\mathcal H_n^{\sUL} [\tau]=1$, the two CSI indicators and their associated \emph{time stamp} $t_n[\tau]$, together constitute a \emph{CSI sample}.
The CSI samples for all $F_n$ frames in the tracking period constitute a \emph{CSI sequence}.
For Band $n$, the CSI sequence is denoted by $\mathcal I_n$ and expressed as:
\beq
\mathcal I_n = \seql (
t_n[\tau], 
\bm H^{\sDT}_n [\tau], 
\boldsymbol{\mathsf{H}}^{\sRIS}_n [\tau]
) 
\!~| \mathcal H_n^{\sUL} [\tau]=1\seqr_{\tau}.
\eeq
Moreover, the CSI sequences of all the $N_{\sB}$ bands are combined as a single sequence denoted by $\mathcal J = \seql \mathcal I_n\seqr_n$ and referred to as an \emph{MB CSI sequence}.

Finally, for the BS to estimate UE track, it employs tracking function $\bm f$ to process $\mathcal J$ and obtains an estimated UE track denoted by sequence ${\mathcal T_{\bm f}}$, which can be expressed as 
\beq
\label{equ: sequence user estimated position}
{\mathcal T}_{\bm f} = \seql  (t_n[\tau], {\bm p}_{\bm f,n} [\tau]) | \mathcal H_n^{\sUL} [\tau]=1\seqr_{n,\tau} = \bm f(\mathcal J),
\eeq 
where ${\bm p}_{\bm f, n}[\tau]\!\in\!\mathbb R^3$ denotes the estimated position of the UE at Frame $\tau$ of Band $n$.\!\footnote{Though interpolation and filtering methods can further extend ${\mathcal T}_{\bm f}$ to a continuous UE track, they are incremental and orthogonal to the core design and optimization of the UE tracking function and thus not considered here.}

\section{\nname\ Framework for Solving Tracking Error Minimization Problem}
\label{sec: prob formulation}
In this section, we propose the \nname\ framework for UE tracking to minimize tracking errors in the considered RIS-aided MB ISAC system.
We first formulate the tracking error minimization problem and analyze its critical challenges.
Then, we describe how \nname\ handles these challenges.
\subsection{Problem Formulation of Tracking Error Minimization}
\label{ssec:formul_track_min}
We focus on minimizing the tracking error of $\bm f$ for the target deployment environment of the system.
Specifically, we evaluate the tracking error by the mean absolute error~(MAE), i.e., mean $\ell^1$-norm distance between the estimated and ground-truth positions within UE tracks, which is referred to as the \emph{axial tracking error} since it evaluates the mean absolute errors along the 3D axes.\!\footnote{\rev{Compared to mean squared error~(MSE) and Euclidean norm error, MAE weighs less on large outlier errors caused by large channel fading along certain directions. This prevents the learning of $\bm f$ in other directions from encountering bottlenecks.}}
Besides, we model the target deployment environment as a \emph{domain}, referred to as the \emph{target domain}, which is characterized by the joint probability distribution of MB CSI sequence $\mathcal J$ and its corresponding ground-truth UE track $\mathcal T=\seql(t_n[\tau], \revs{\bm p_n[\tau]}) | \mathcal H_n^{\sUL} [\tau]=1\seqr_{n, \tau}$, denoted by $\oPr^{\stgt}: (\mathcal J, \mathcal T)\rightarrow [0,1]$.
Therefore, the tracking error minimization problem can be formulated as:
\begin{align}
\label{equ: P0}
\text{(P0): }\min_{\bm f}~ %
\varepsilon^{\stgt}(\bm f) \! =\! 
& \mathop{\mathbb E}\limits_{(\mathcal J, \mathcal T)\sim \oPr^{\stgt}} 
\!\Big( \! \frac{1}{|\mathcal J|} \sum_{n=1}^{N_{\sB}} \sum_{\tau=1}^{F_n}   \mathcal H_n^{\sUL} [\tau] \nonumber \\
& \cdot \frac{1}{3}\big\| \revs{\bm p_n[\tau]} -  {\bm p}_{\bm f,n}[\tau] \big\|_1\Big),
\end{align}
where $|\mathcal J| = \sum_{n} \sum_{\tau}\mathcal H^{\sUL}_n[\tau]$ is the total number of CSI samples (or equivalently, the number of positions considered in the UE track), and factor $1/3$ accounts for the averaging across the 3D axes.
Moreover, we refer to $\varepsilon^{\stgt}(\bm f)$ as the \emph{expected axial tracking error for the target domain}.
Nevertheless, solving (P0) faces three crucial challenges:

\textbf{Diverse Multi-modal CSI Indicators}: In~(P0), $\bm f$ needs to handle the MB CSI sequences that comprise CSI indicators with diverse modalities.
Specifically, in each band, DT-CSI and RIS-CSI have different dimensions and statistical characteristics as they represent the channel gains for distinct paths. 
Moreover, the CSI indicators of different bands also have different modalities due to the distinct signal propagation characteristics of each band. 
Consequently, there are $2N_{\sB}$ data modalities in $\mathcal J$, which are challenging for $\bm f$ to jointly utilize in an efficient manner.

\textbf{Irregular and Asynchronous MB CSI Sequences}: 
In $\mathcal J$, CSI sequences $\mathcal I_{1},\cpdots,\mathcal I_{N_{\sB}}$ are composed of CSI samples that are irregularly spaced in time due to the irregularity of the realistic UL traffic~\cite{ChenCOMM23_ISACoT} and the UE scheduling of the BS.
Moreover, the UL traffic of a UE and the scheduling of the BS is generally inconsistent across multiple bands, resulting in the MB CSI sequences being asynchronous.
Such irregularity and asynchrony of the MB CSI sequences are hard to handle~\cite{Narayan2021ICLR_Multi,Zhang2022ICLR_Graph}, making it challenging for $\bm f$ to process CSI sequences and achieve effective mutual complementation for the positional information contained in MB CSI.

\textbf{Sparse Labeled Data for Learning}: 
To evaluate the objective function in~(P0), it requires $\oPr^{\stgt}$ to be known, or equivalently, sufficient labeled data for $\oPr^{\stgt}$ to be available.
Here, the \emph{labeled data} comprise \emph{data}, i.e., the MB CSI sequences, and \emph{labels}, the ground-truth UE tracks.
However, in the target deployment environment, it is impractical to obtain a sufficient amount of labeled data since it requires cooperative UEs with self-positioning capabilities to travel throughout the ROI, which is prohibitively time-consuming and interferes with other normal UEs.
In general, only a dataset of sparse labeled data can be obtained with a few cooperative UEs traveling in a fraction of the ROI.
This lack of labeled data for $\oPr^{\stgt}$ makes $\varepsilon^{\stgt}(\bm f)$ intractable and~(P0) challenging to solve.

\subsection{\nname\ Framework}
\label{ssec:framework}
To handle the first two challenges of (P0), \nname\ represents $\bm f$ with a hierarchical architecture:
For each band, the multi-modal CSI indicators due to the RIS-aided CSI collection are handled by modal-specific CSI encoders and jointly aggregated into feature vectors.
\revs{Subsequently, the feature vectors of MB CSI are integrated with features of their irregular and asynchronous sampling time and projected into a common feature space, where they can mutually complement based on their correlation.
Thanks to the integrated time features, the correlation of feature vectors can process their intrinsic positional information considering their time relevance, allowing the mutual complementation of positional information of MB CSI seqeunces in spite of their asynchrony.}
Then, to tackle the sparsity of labeled data, \nname\ converts the objective function of~(P0) through cross-domain learning based on the \blue{domain adaptation~(DA) theorem~\cite{Ben_David2010ATheory}}.
\subsubsection{Hierarchical Architecture of Tracking Function}
\label{s3ec: model of f}

In \nname, $\bm f$ is first decomposed into a composite function comprising two parts: a \emph{feature encoder} $\bm h$ and a \emph{position regressor} $\bm \xi$, i.e., $\bm f = \bm \xi \circ \bm h$.
To explicitly handle multi-modal CSI indicators and irregular and asynchronous MB-CSI sequences, \nname\ decomposes $\bm h$ into two-level modules: $N_{\sB}$ \emph{frame-level encoders} and a \emph{sequence-level encoder}.

For each CSI sample, frame-level encoders process the DT-CSI and RIS-CSI in a modal-specific manner to extract features of CSI indicators and then aggregate them as \emph{CSI feature vectors}.
More specifically, each frame-level encoder contains three functions: $\bm \psi_n^{\sDT}$, $\bm\psi_n^{\sRIS}$, and $\bm \psi_n^{\sCSI}$.
Here, $ \bm \psi_n^{\sDT}: \mathbb C^{Q_n\times N_{\sSC,n}}\rightarrow \mathbb R^{L^{\sDT}_n}$ and 
$\bm\psi_n^{\sRIS}: \mathbb C^{Q_n\times N_{\sSC,n}\times M_n} \rightarrow \mathbb R^{L^{\sRIS}_n}$ 
are modal-specific encoders for DT-CSI and RIS-CSI of Band~$n$, with $L^{\sDT}_n$ and $L^{\sRIS}_n$ being the lengths of the resulting feature vectors for DT-CSI and RIS-CSI, respectively.
Therefore, for the CSI sample of Frame $\tau$ in $\mathcal I_n$, the obtained DT-CSI and RIS-CSI feature vectors can be expressed as:
\begin{align}
\bm x_n^{\sDT} [\tau] = \bm \psi_n^{\sDT} (\bm H^{\sDT}_n [\tau]), ~
\bm x_n^{\sRIS} [\tau] = \bm \psi_n^{\sRIS} (\boldsymbol{\mathsf{H}}^{\sRIS}_n [\tau]). \nonumber 
\end{align}
Besides, $\bm \psi_n^{\sCSI}:{\mathbb R}^{L^{\sDT}_n+L^{\sRIS}_n}\!\rightarrow\!\mathbb R^{L}$ aggregates $\bm x_n^{\sDT} [\tau]$ and $\bm x_n^{\sRIS} [\tau]$, mapping the concatenated vector to a CSI feature vector of length $L$, i.e., $\bm x^{\sCSI}_n[\tau]=\bm\psi^{\sCSI}_n(\bm x_n^{\sDT} [\tau]\oplus \bm x_n^{\sRIS} [\tau])$.

Subsequently, in the sequence-level encoder, the CSI feature vectors are first tagged with its time features, enabling the extraction of their temporal relevance and thus facilitating their mutual complementation.
This is achieved by a \emph{time encoder} $\bm \psi^{\sTM}:\mathbb R\!\times\! \mathbb R^{L}\rightarrow \mathbb R^{L}$ encoding time stamp $t_n[\tau]\in\mathbb R$ into the CSI feature vector, which is represented as $\bm x_n[\tau]=\bm \psi^{\sTM}_n(t_n[\tau], \bm x^{\sCSI}_n[\tau])$.
Following that, the sequences of $\bm x_n[\tau]$ for the $N_{\sB}$ bands are jointly input into a sequence-to-sequence encoder denoted by $\bm \chi: \mathbb R^{|\mathcal J|\times L}\rightarrow \mathbb R^{|\mathcal J|\times L}$, i.e.,
\beq
\label{equ: seq-level encoder}
\seql \bm  y_n[\tau] | \mathcal H_n^{\sUL} [\tau] \!=\! 1\seqr_{n, \tau} 
\!= \!
\bm\chi(\seql \bm  x_n[\tau] | \mathcal H_n^{\sUL} [\tau]\!=\! 1\seqr_{n, \tau} ).
\eeq
where $\seql \bm  y_n[\tau] | \mathcal H_n^{\sUL} [\tau] = 1\seqr_{n,\tau}$ is referred to as the sequence of \emph{positional feature vectors}.

\rev{Intuitively, in~\eqref{equ: seq-level encoder}, the input CSI feature vectors are projections of the time-encoded CSI samples of multiple bands onto a common feature space.
In this feature space, $\bm \chi$ allows the positional information in the input vectors to mutually complement based on their correlation.
By this means, the difficulties of characterizing temporal variations of irregularly-spaced CSI samples and aligning asynchronous MB CSI sequences are both avoided.}
Here, the mutual complementation encompasses three aspects:~a) The CSI feature vectors suffered from non-LoS channel conditions are complemented by those collected under LoS conditions;~b) The noise and strong reflection interference are mitigated by extracting the correlated features of neighboring CSI samples;~c) Jointly using the CSI feature vectors across multiple bands increases the number of available CSI samples to estimate the UE track and extends the effective bandwidth, improving the spatial resolution.

\rev{As the CSI feature vectors from $N_{\sB}$ bands are mutually complemented by $\bm \chi$, the joint utilization of MB CSI is achieved.
Therefore, the resulting positional feature vectors can be processed individually and uniformly by a position regressor, requiring no additional combining operations.}
We denote the position regressor by $\bm \xi:\mathbb R^{L}\rightarrow \mathbb R^{3}$, and the estimated UE track can be expressed as:
\beq
\label{equ:regressor_to_p}
{\bm p}_{\bm f, n}[\tau] = \bm\xi(\bm y_n[\tau]),
\eeq
$\forall n\in\{1,\cpdots,\!N_{\sB}\}$, $\tau\in\{1,...,F_n\}$, given $\mathcal H_n^{\sUL} [\tau]=1$.
Combining the time stamp with each CSI sample, \nname\ obtains the estimated track of the UE, i.e., ${\mathcal T}_{\bm f}$ in~\eqref{equ: sequence user estimated position}.
In summary, $\bm f$ can be represented by:
\beq
\label{equ: f decompose}
\bm f = \bm \xi \circ \bm h= \bm \xi \circ \bm \chi \circ \seql \bm \psi_n^{\sTM} \circ \bm \psi_n^{\sCSI}\circ (\bm \psi^{\sDT}_n \oplus \bm \psi_n^{\sRIS})\seqr_n .
\eeq
The hierarchical architecture in~\eqref{equ: f decompose} simplifies the design and optimization of $\bm f$, allowing them to be divided and conquered.
By representing $\bm f$ with the hierarchical architecture, \nname\ decomposes the functionalities of $\bm f$ into multiple modules, explicitly specifying how multi-modal CSI indicators and their irregularly and asynchronous sequences are handled.

\subsubsection{DA based Cross-domain Learning}
\label{s3ec: converting da prob}
\nname\ handles the challenge of evaluating the objective function in (P0) through cross-domain learning based on the DA~\cite{Ben_David2010ATheory}.
The cross-domain learning leverages a labeled dataset for a known domain, namely a \emph{source domain} denoted by $\oPr^{\ssrc}$~($\oPr^{\ssrc}\neq\oPr^{\stgt}$), to help optimize $\bm f$ over the sparse labeled dataset for $\oPr^{\stgt}$.
Similar to $\oPr^{\stgt}$, $\oPr^{\ssrc}$ is the probability distribution for the data and labels in a \emph{source deployment environment}, e.g., an experimental site, where sufficient ground-truth labels can be obtained.
\blue{The source and target deployment environments are expected to be ``distinct yet similar''~\cite{Zhang2019ICML_Bridging}}.
For instance, they can have different locations of the RISs, types of UEs, channel conditions, etc, while the relationships between CSI samples and positional features in both environments are similar.

We denote the sets of labeled data for $\oPr^{\ssrc}$ and $\oPr^{\stgt}$ by $\mathcal D^{\ssrc}$ and $\mathcal D^{\stgt}$, respectively, noting that $|\mathcal D^{\ssrc}|\gg |\mathcal D^{\stgt}|$ due to sparsity of labeled data for $\varGamma^{\stgt}$.
Besides, we denote the sets of unlabeled data for $\oPr^{\stgt}$ and $\oPr^{\ssrc}$ by $\bar{\mathcal D}^{\stgt}= \{ \mathcal J| \mathcal J\sim \oPr^{\stgt}_{\mathcal J}\}$ and $\bar{\mathcal D}^{\ssrc}= \{ \mathcal J| \mathcal J\sim \oPr^{\ssrc}_{\mathcal J}\}$, \rev{using short bars over datasets to indicate unlabeled datasets.}
Here, $\oPr^{\stgt}_{\mathcal J}$ and $\oPr^{\ssrc}_{\mathcal J}$ represent the marginal distributions of $\mathcal J$ for $\oPr^{\stgt}$ and $\oPr^{\ssrc}$, respectively. 
Moreover, we assume $|\bar{\mathcal D}^{\stgt}| = |\bar{\mathcal D}^{\ssrc}| \gg |\mathcal D^{\stgt}|$.
Then, based on DA theorems in~\cite{Ben_David2010ATheory,Li2021CVPR_Learning,Zhao2018NIPS_Adversarial}, two upper bounds related to $\varepsilon^{\stgt}(\bm f)$ can be derived in Lemma~\ref{prop: two bounds}.

\begin{lemma}
\label{prop: two bounds}
Index the axes of 3D coordinate by $a=1,2,3$, and denote the expected axial tracking error in (P0) of Axis $a$ for the target domain by $\varepsilon_{a}^{\stgt}(f_a)$, with $f_a$ being the part of $\bm f$ for Axis $a$.
Additionally, represent the empirical evaluation by hats over the functions, e.g., $\hat{\varepsilon}_a^{\stgt}(f_a)$ represents the empirical evaluation of $\varepsilon_a^{\stgt}(f_a)$ over $\mathcal D^{\stgt}$.
\rev{Moreover, denote by $\mathcal X_a$ the feasible set of function ${f_a}$.}
Then, with probability $1-\delta$~($\delta\in(0,1)$), $\varepsilon^{\stgt}_{a}(f_a)$ has the following two upper bounds:
\begin{align}
\label{equ: up bound 1}
\varepsilon_{a}^{\stgt}(f_a) 
& \leq 
\hat{\varepsilon}_{a}^{\ssrc}(f_a) + {S_a}\hat{d}_{\tilde{\mathcal X}_a}(\bar{\mathcal D}^{\ssrc}, \bar{\mathcal D}^{\stgt})  \\
&
+ 2 S_a \sqrt{\frac{2\mathrm{Pdim}(\mathcal X_a)\log(2|\bar{\mathcal D}^{\ssrc}|+2)+2\log(8/\delta)}{|\bar{\mathcal D}^{\ssrc}|}} \nonumber\\[-.2em]
& 
+ 4 S_a\sqrt{\frac{2\mathrm{Pdim}(\mathcal X_a)\log(2|\bar{\mathcal D}^{\stgt}|)+\log(2/\delta)}{|\bar{\mathcal D}^{\stgt}|}} + \lambda^*_a,  \nonumber\\
\label{equ: up bound 2}
\varepsilon_{a}^{\stgt}(f_a) 
& \leq 
\hat{\varepsilon}_{a}^{\stgt}(f_a)  \\
& +2 S_a \sqrt{\frac{2\mathrm{Pdim}(\mathcal X_a)\log(2|{\mathcal D^{\stgt}}|+2)+2\log(8/\delta)}{|{{\mathcal D}^{\stgt}}|}},\nonumber
\end{align}
where $S_a$ is the side length of the ROI along Axis $a$,
$\mathrm{Pdim}(\mathcal X_a)$ represents the pseudo-dimension of the functions in $\mathcal X_a$~\cite{Li2021CVPR_Learning},
and $\lambda^*_a = \min_{f_a} (\varepsilon_{a}^{\stgt}(f_a) + \varepsilon_{a}^{\ssrc}(f_a))$ is the minimum expected axial tracking error for the source and target domains of Axis $a$.
Besides, $\hat{d}_{\tilde{\mathcal X}_a}(\bar{\mathcal D}^{\ssrc}, \bar{\mathcal D}^{\stgt})$ represents the $\mathcal H$-divergence measure between sets $\bar{\mathcal D}^{\ssrc}$ and $\bar{\mathcal D}^{\stgt}$ and is determined by 
\beq
\label{equ: divergence}
\hat{d}_{\tilde{\mathcal X}_a}(\bar{\mathcal D}^{\ssrc}, \!\bar{\mathcal D}^{\stgt}) \!=\! \sup_{\zeta\in\tilde{\mathcal X}_a}\!\Big|\frac{1}{|\bar{\mathcal D}^{\stgt}|}\!\sum_{\mathcal J\in\bar{\mathcal D}^{\stgt}}\hspace{-.5em}\zeta(\mathcal J)-\frac{1}{|\bar{\mathcal D}^{\ssrc}|}\!\sum_{\mathcal J\in\bar{\mathcal D}^{\ssrc}}\hspace{-.5em}\zeta(\mathcal J)\Big|,
\eeq
where \rev{$\tilde{\mathcal X}_a \!=\! \{ \zeta\! =\! \mathbb I(\|{f}_a(\mathcal J) \! - \! {f}_a'(\mathcal J)\|_1\!>\!S_a\cdot\iota)~|~{f}_a, {f}_a' \!\in\! \mathcal X_a, \iota\!\in\![0,1]\}$} is the set of threshold functions induced from $\mathcal X_a$.
\end{lemma}
\begin{IEEEproof}
Inequality~\eqref{equ: up bound 1} is derived by jointly using~\cite[Thm.~2 and Lemma~5]{Ben_David2010ATheory} and the proof of \cite[Lemma~2]{Ben_David2010ATheory} and substituting the $\mathcal H\Delta \mathcal H$-divergence term with~\eqref{equ: divergence} based on~\cite[Thm.~4.2]{Li2021CVPR_Learning}.
Inequality~\eqref{equ: up bound 2} is derived by using \cite[Lemma~5]{Ben_David2010ATheory} and the proof of \cite[Thm.~3]{Ben_David2010ATheory}.
\end{IEEEproof}

Based on~\eqref{equ: up bound 1}, an available approach to solve (P0) is by converting its objective to minimizing the sum of the empirical axial tracking error for the source domain and the divergence between the source and target domains.
Alternatively, based on~\eqref{equ: up bound 2}, another possible approach is to convert the objective of (P0) to minimizing the empirical axial tracking error for the target domain over $\mathcal D^{\stgt}$.

However, a closer examination of~\eqref{equ: up bound 1} and~\eqref{equ: up bound 2} reveals the deficiency in the two approaches.
Firstly, the tightness of the upper bound in~\eqref{equ: up bound 1} relies on the assumption that $\lambda^*_a$ is small, i.e., a tracking function exists that works well for \emph{both} the source and target domains.
If this assumption does not hold, the large $\lambda^*_a$ will dominate the empirical loss for the source domain and the divergence term.
Secondly, for the upper bound in~\eqref{equ: up bound 2}, the sparsity of labeled data for the target domain can lead to significant discrepancies between the empirical and the expected axial tracking errors, especially for a complex function as $\bm f$ that is prone to over-fitting.
This is explicitly indicated by the large second term on the right-hand side of~\eqref{equ: up bound 2} in the case of small $|{\mathcal D^{\stgt}}|$ and large $\mathrm{Pdim}(\mathcal X_a)$.

To overcome the deficiency associated with relying solely on either approach, we propose to use them combinedly while leveraging the hierarchical architecture of $\bm f$ in \nname.
As a result, we derive a new upper bound for the objective function of (P0) in Proposition~\ref{prop: new upper bound}.

\begin{proposition}
\label{prop: new upper bound}
\rev{Assume that the position regressors for 3D axes, i.e., ${\xi}_1$, ${\xi}_2$, and ${\xi}_3$, are of the same architecture $\xi$, and denote the feasible set of $\xi$ by $\mathcal Z$.}
Denote by $\mathcal F^{\ssrc}(\bm h)$ and $\mathcal F^{\stgt}(\bm h)$ the sets of positional feature vectors obtained by encoding data in $\bar{\mathcal D}^{\ssrc}$ and $\bar{\mathcal D}^{\stgt}$ with $\bm h$.
Then, with probability larger than $(1-\delta)$~($\delta\!\in\!(0,1)$), $\varepsilon^{\stgt}(\bm f)$ is upper-bounded by
\begin{align}
\label{equ: new up bound}
\varepsilon^{\stgt}(\bm f) 
& \leq
~\hat{\varepsilon}^{\stgt}(\bm \xi\circ \bm h) \nonumber\\
& + 2 S \sqrt{\frac{2 \mathrm{Pdim}(\mathcal Z)\log(2|{\mathcal D^{\stgt}}|+2)+2\log(8/\delta)}{|{\mathcal D^{\stgt}}|}} \nonumber\\[-.3em]
& + \min_{\bm \xi'} \hat{\varepsilon}^{\ssrc}(\bm \xi' \!\circ\! \bm h) + S \cdot \hat{d}_{\tilde{\mathcal Z}}(\mathcal F^{\ssrc}(\bm h), \mathcal F^{\stgt}(\bm h)),
\end{align}
where ${S} = \sum_{a=1}^3S_a/3$, and $\tilde{\mathcal Z}$ and $\hat{d}_{\tilde{\mathcal Z}}(\cdot)$ are defined similarly to $\tilde{\mathcal X}_a$ and $\hat{d}_{\tilde{\mathcal X}_a}(\cdot)$ in Lemma~\ref{prop: two bounds}.
\rev{The equality holds with probability $1$ when $\bm h$ satisfies $\min_{\bm \xi'} \hat{\varepsilon}^{\ssrc}(\bm \xi' \!\circ \bm h)=0$ and $\hat{d}_{\tilde{\mathcal Z}}(\mathcal F^{\ssrc}(\bm h), \mathcal F^{\stgt}(\bm h))=0$, and $|\mathcal D^{\stgt}|\rightarrow \infty$.}
\end{proposition}
\begin{IEEEproof}
\rev{For a given $\bm h$, adapt inequalities~\eqref{equ: up bound 1} and~\eqref{equ: up bound 2} to $\xi_a$~($a\!=\! 1,2,3$), respectively, and average the results for $a\!=\!1,2,3$. 
Then, for the adapted~\eqref{equ: up bound 1}, taking minimization on both sides w.r.t. the position regressor, resulting in the first and second terms on the right-hand side being $ \min_{\bm \xi'} \hat{\varepsilon}^{\ssrc}(\bm \xi' \!\circ \bm h)$ and $S \cdot \hat{d}_{\tilde{\mathcal Z}}(\mathcal F^{\ssrc}(\bm h), \mathcal F^{\stgt}(\bm h))$, respectively.
Since these two terms are non-negative, they can be added to the right-hand side of the adapted~\eqref{equ: up bound 2}, and then~\eqref{equ: new up bound} is derived.}
\end{IEEEproof}

The upper bound in \eqref{equ: new up bound} has three advantages over the ones in~Lemma~\ref{prop: two bounds}:
\rev{\emph{Firstly}, it is based on the adapted~\eqref{equ: up bound 2} which focuses on the positional regressor rather than the tracking function as in~\eqref{equ: up bound 2}, leading to a much tighter bound due to $\mathrm{Pdim}(\mathcal Z)\!\ll\!\mathrm{Pdim}(\mathcal X_a)$.
\emph{Secondly}, by incorporating $ \min_{\bm \xi'} \hat{\varepsilon}^{\ssrc}(\bm \xi' \!\circ\! \bm h)$ and $S \cdot \hat{d}_{\tilde{\mathcal Z}}(\mathcal F^{\ssrc}(\bm h), \mathcal F^{\stgt}(\bm h))$ derived from the adapted~\eqref{equ: up bound 1}, it efficiently exploits the large sets of labeled and unlabeled data for the source and target domains, which enables effective evaluations on the general capability of $\bm h$ in extracting positional features from MB CSI sequences.}
\emph{Thirdly}, by removing $\lambda_a^*$ and allowing the source and target domains to have separated position regressors, i.e., $\bm\xi'$ being independent of $\bm \xi$, its tightness does not suffer from the nonexistence of an effective tracking function working well for both the source and target domains.
\rev{These three advantages indicate that the upper bound in~\eqref{equ: new up bound} is more efficient for the tracking function optimization than those in Lemma~\ref{prop: two bounds}.}

\rev{Therefore, \name\ adopts the upper bound in~\eqref{equ: new up bound} as a surrogate function for the original objective function $\varepsilon^{\stgt}(\bm f)$, handling the challenge of evaluating $\varepsilon^{\stgt}(\bm f)$.
Due to the upper bound being closely related to $\varepsilon^{\stgt}(\bm f)$, minimizing the upper bound can reduce $\varepsilon^{\stgt}(\bm f)$ effectively.}
Omit the constant term in the upper bound, and the converted optimization problem can be formulated as:
\begin{align}
\text{(P1)}~\bm f^* =\mathop{\arg}\limits_{\bm \xi \circ \bm h} \min_{\bm \xi, \bm \xi', \bm h} & ~\hat{\varepsilon}^{\stgt}(\bm \xi \circ\bm h) + \hat{\varepsilon}^{\ssrc}(\bm \xi'\circ\bm h) \nonumber\\[-.3em]
& + S\cdot \hat{d}_{\tilde{\mathcal Z}}(\mathcal F^{\ssrc}(\bm h), \mathcal F^{\stgt}(\bm h)).
\end{align}

\rev{By converting~(P0) to~(P1), \name\ handles the challenge of sparse labeled data and allows the optimization to have sufficient training data for the convergence of $\bm f$.
Specifically, on the right-hand side of (P1), the second and third terms are evaluated with the large dataset for the source domain and the large unlabeled dataset for the target domain.
Therefore, by optimizing the second and third terms, $\bm h$ is ensured to extract effective positional feature vectors from MB CSI sequences for both the source and target domains.
Though the first term is evaluated with the sparse labeled data, it is sufficient for the convergence of $\bm \xi$. 
This is because $\bm \xi$ learns a much less complex mapping than $\bm h$ and has a much simpler architecture, allowing it to be trained with substantially fewer data.}

In summary, the hierarchical architecture of $\bm f$ and the converted problem~(P1) constitute the \nname\ framework that solves the tracking error minimization problem~(P0).

\section{Algorithm Design under \nname\ Framework}
\label{sec: alg design}

We design an efficient algorithm to solve the tracking error minimization problem under the \nname\ framework.
In subsection~A, we design $\bm f$ as a transformer-based DNN\footnote{
	For presentation and implementation clarity, our algorithm comprises standard DNN blocks. 
\rev{Other algorithms with more sophisticated DNNs may improve the tracking performance while incurring higher computational complexity. Exploring such tradeoffs and pursuing the optimality of the algorithm is beyond the scope of this paper and thus left for future work.}
	} illustrated in Fig.~\ref{fig: feature extractor}, following the hierarchical architecture proposed in Sec.~\ref{s3ec: model of f}.
\rev{In subsection~B, with the help of adversarial learning techniques, we design an algorithm that jointly trains the frame-level and sequence-level encoders and the position regressor in Fig.~\ref{fig: feature extractor} to solve~(P1).}

\subsection{Transformer-based Tracking Function}
\label{s2ec: MM CB Transformer}

\begin{figure}[t] 
\centering
\includegraphics[width=1\linewidth]{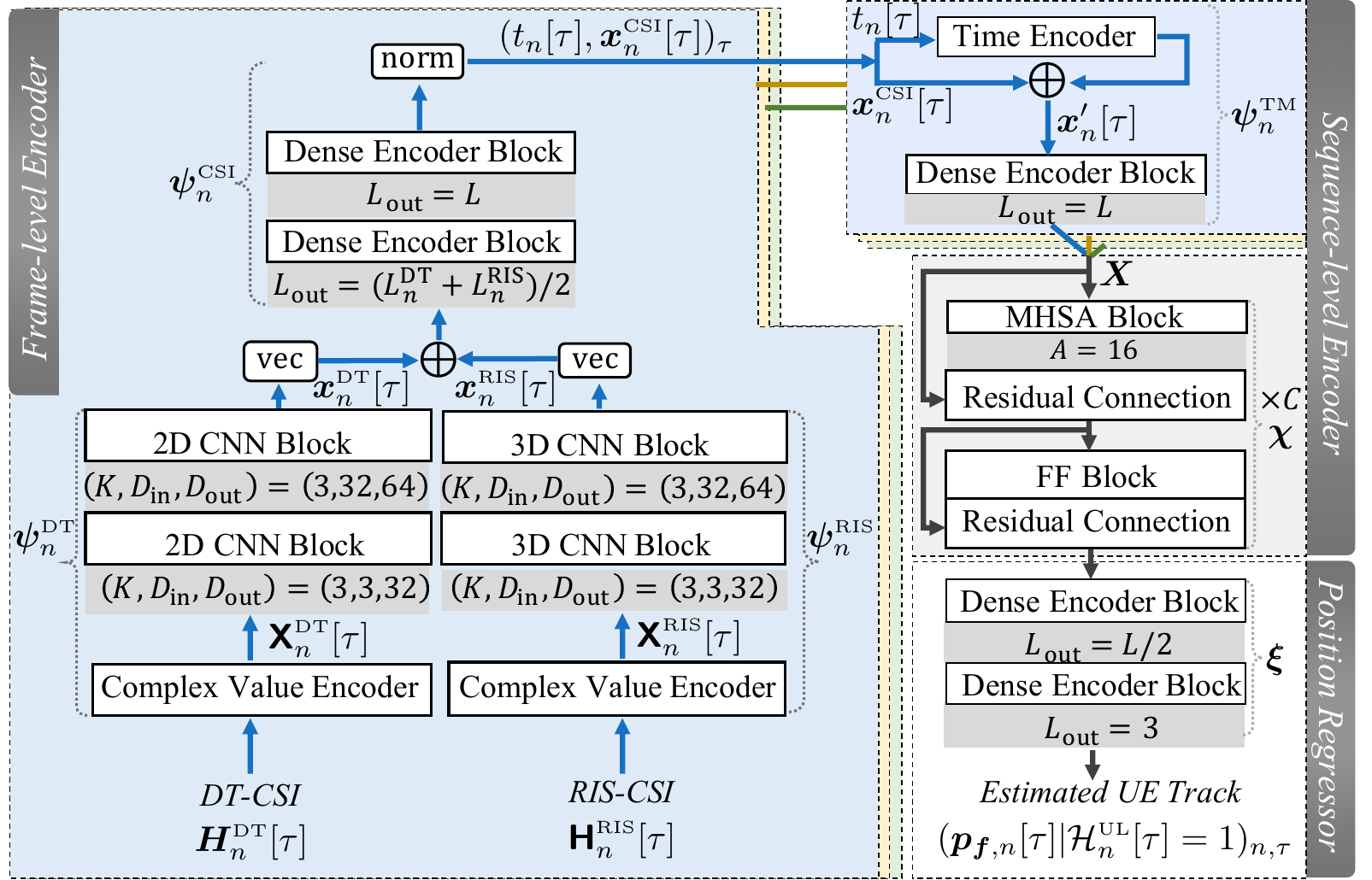}
  \vspace{-1.em}
  \caption{\revs{Transformer-based tracking function $\bm f$ based on the hierarchical architecture proposed in the \nname\ framework.}}
  \label{fig: feature extractor}
\vspace{-.5em}
\end{figure}

According to Sec.~\ref{s3ec: model of f}, $\bm f$ comprises $N_{\sB}$ frame-level encoders, a sequence-level encoder, and a position regressor.
In light of the recent prominent success of transformers in handling sequence data, we design $\bm f$ to be featured by a transformer-based sequence-level encoder that utilizes the multi-head self-attention~(MHSA) mechanism.
The detailed design of $\bm f$ is depicted in Fig.~\ref{fig: feature extractor}.

\subsubsection{Frame-level Encoders}
\label{s3ec: frame-level encoder}

For each Band $n$, the frame-level encoder, i.e., $\bm \psi_n^{\sCSI}\circ (\bm \psi^{\sDT}_n \oplus \bm \psi_n^{\sRIS})$, comprises complex value encoders, 2D and 3D convolutional neural network~(CNN) blocks, and dense encoder blocks.
The complex value encoders convert the complex-valued CSI into its real-valued form.
The CNN blocks extract local patterns and regional information and reduce the data dimensionality of CSI indicators.
The dense encoder blocks further refine and compress the concatenated feature vectors of DT-CSI and RIS-CSI by recognizing the interrelationship among their elements. 
Further details on these key component modules are provided below.

\textbf{Complex Value Encoder}:
As DNNs are more efficient to handle real-valued input than complex-valued ones, we use complex value encoders to represent each $x\in \mathbb C$ as a real-valued 3-tuple $(|x|, \cos(\angle x), \sin(\angle x))$. 
\rev{By encoding $\angle x$ with cosine and sine functions, the encoder preserves the cyclicity of $\angle x$ and avoids the discontinuity between $0$ and $2\pi$, which helps $\bm f$ capture actual phase difference between complex values.}
In addition, to remove the dependency on specific value ranges, a max-min normalization is adopted to force the value range to be $[-1,1]$.
The encoded $\bm H^{\sDT}_n [\tau]$ 
and $\boldsymbol{\mathsf{H}}^{\sRIS}_n [\tau]$ are denoted by 
${ \bm{\mathsf{X}} }^{\sDT}_n[\tau] \!\in\!\mathbb R^{Q_n\times N_{\sSC,n}\times 3}$ and 
${ \bm{\mathsf{X}}}^{\sRIS}_n[\tau]\! \in\!\mathbb R^{Q_n\times N_{\sSC,n}\times M_n\times 3}$, respectively.
As per the conventions, we refer to the last dimension as the \emph{channel dimension}.

\textbf{2D/3D CNN Block}:
The 2D and 3D CNN blocks handle $ \bm{\mathsf{X}}^{\sDT}_n[\tau] $ and $ \bm{\mathsf{X}}^{\sRIS}_n[\tau]$, respectively.
Each block comprises a CNN layer, an activation layer, a normalization layer, and an average pooling layer.
Here, we describe a 2D CNN block as an example.
\revs{For each channel, the CNN layer convolves the input multi-channeled 2D data with a 2D kernel of sidelength $K$, extracting and encoding local and regional features of the input.
Following that, the activation layer introduces non-linearity into the DNN by using the ReLU activation funciton $\relu(x) = \max(x, 0)$.
Then, the normalization layer normalizes the resulting elements to have a zero mean and unit variance for generalizability. 
Finally, an average pooling layer compresses its input along the first two dimensions by averaging neighboring four elements in each channel.}

\textbf{Dense Encoder Block}: Each dense encoder block comprises a densely connected layer and an activation layer.
Given output length $L_{\sOut}$, it maps input $\bm x_{\sIn}$ of length $L_{\sIn}$ to $\bm x_{\sOut} = \relu(\bm W \bm x_{\sIn} + \bm b)$, where $\bm W\!\in\! \mathbb R^{L_{\sOut}\times L_{\sIn}}$ and $\bm b \!\in\! \mathbb R^{L_{\sOut}}$ are its trainable weight and bias parameters.

\subsubsection{Sequence-level Encoder}
\label{s3ec: ibc module}

The sequence-level encoder is composed of time encoders $\bm\psi^{\sTM}_1,\cpdots,\! \bm\psi_{N_{\sB}}^{\sTM}$ and sequence-to-sequence encoder $\bm \chi$.
As transformers are universal approximators for sequence-to-sequence functions~\cite{Yun2020ICLR_Are}, we design a transformer-based sequence-level encoder, utilizing the core MHSA mechanism of transformers to achieve mutual complementation of MB CSI sequences.
More specifically, $\bm \chi$ is composed of $C$ cascaded MHSA blocks and feed-forward~(FF) blocks, with each block followed by a \emph{residual connection} to alleviate the vanishing gradient problem of DNNs~\cite{He2016_CVPR}.

\textbf{Time Encoder}:
To capture the temporal periodicity, we employ a series of periodic trigonometric functions to encode each scaler time stamp $t_n[\tau]\in\mathbb R$ to vector $\bm x^{\sTM}_n[\tau]\in\mathbb R^{L_{\sTM}}$:
\beq
\left[ \bm x^{\sTM}_n[\tau] \right]_d 
= \begin{cases}
 	\sin(t_n[\tau]/ T_{\sTP}^{d/L_{\sTM}}),&\text{if $d$ is even,}\\[-.0em]
 	\cos(t_n[\tau]/ T_{\sTP}^{(d-1)/L_{\sTM}}),&\text{otherwise,}
 \end{cases}\nonumber
\eeq
$d = 1,\cpdots, \! L_{\sTM}$.
As shown in Fig.~\ref{fig: feature extractor}, $\bm x^{\sTM}_n[\tau]$ is concatenated with $\bm x^{\sCSI}_n[\tau]$, and the result, i.e., $\bm x_n'[\tau] = \bm x_n^{\sCSI}[\tau] \oplus \bm x_n^{\sTM}[\tau]$ of length $L + L_{\sTM}$, is projected to $\bm x_n[\tau]\in\mathbb R^{L}$ by an additional dense encoder block to keep the length of the feature vector unchanged. 
In addition, to facilitate further processing, sequence $\seql \bm x_n[\tau] | \mathcal H_n^{\sUL} [\tau_{n}]=1 \seqr_{n, \tau}$ is reshaped into matrix $\bm X$ with size $|\mathcal J| \times L$.

\textbf{MHSA Block}:
To describe the MHSA block clearly, we begin by introducing the \emph{attention mechanism}.
The attention mechanism maps input matrices comprising \emph{query} and \emph{key} vectors to an output matrix comprising weighted \emph{value} vectors, where value vectors are linear encodings of the key vectors~\cite{vaswani2017attention}.
More specifically, given matrix of query vectors $\bm X_{\sQ}\in \mathbb R^{N_{\sQ}\times L_{\sQ}}$ and matrix of key vectors $\bm X_{\sK} \in \mathbb R^{N_{\sK}\times L_{\sK}}$, the output matrix is calculated by
\beq
\label{equ: attn mech}
\attn(\bm X_{\sQ}, \!\bm X_{\sK})\! = \!
\softmax\left(\frac{
        (\bm X_{\sQ} \!\bm W_{ \sQ })
        (\bm X_{\sK} \!\bm W_{ \sK })^\tran}
       {\sqrt{L_{\sK}}}
    \right)\!(\bm X_{\sK} \!\bm W_{ \sV }),
\eeq
where $\softmax~(\cdot)$ is a row-wise soft-max function,
$\bm W_{\sQ}\!\in\!\mathbb R^{L_{\sQ}\times L_{\sK} }$, $\bm W_{\sK}\!\in\!\mathbb R^{L_{\sK} \times L_{\sK} }$, and $\bm W_{\sV}\!\in\! \mathbb R^{L_{\sK} \times L_{\sQ}} $ are trainable parameters.
\blue{The result of the soft-max function in~\eqref{equ: attn mech}} is referred to as  \emph{attention scores}, which measure the mutual correlation among the row vectors of $\bm X_{\sQ}$ and $\bm X_{\sK}$.
When $\bm X_{\sQ} \!=\! \bm X_{\sK}$ and denoted by $\bm X_{\sIn}$, $\attn(\bm X_{\sIn}, \!\bm X_{\sIn})$ is referred to as the \emph{self-attention}~(SA) mechanism.
In this case, row vectors of $\bm X_{\sIn}\bm W_{\sQ}$, $\bm X_{\sIn} \bm W_{ \sK }$, and $\bm X_{\sIn} \bm W_{ \sV }$ are linear encodings of the input vectors.
Through the summation of the relevant linear encodings of each input vector weighted by their corresponding attention scores, the SA mechanism enables the information in individual input feature vectors to mutually complement.

As for an MHSA block with a number of $A$ attention heads, the input matrix is first divided along its column dimension into $A$ sub-matrices, and then each sub-matrix is handled by the SA mechanism separately, allowing the mutual correlation among the input vectors to be calculated for different aspects.
\rev{For example, denote the input sub-matrices of an MHSA block by $\bm X_{1},\cpdots, \bm X_{A}$, its output can be expressed as $\bm X_{\sOut} = \attn(\bm X_{1},\bm X_{1})\oplus\cpdots \oplus\attn(\bm X_{ A }, \bm X_{A})$.}

\textbf{FF Block}: 
Following an MHSA block, an FF block handles each row vector of the input matrix by a densely connected layer where both the input and output sizes are $L$.
As a result, the FF block combines and refines the information contained in the elements of each feature vector.

\textbf{Residual Connection}: 
As the number of cascaded blocks increases, the impact of the earlier blocks on the later outputs becomes hard to trace, resulting in the vanishing gradient problem of DNNs~\cite{Hanin2018Which_NIPS}. 
To mitigate this issue, residual connections are employed to create shortcuts for the impact to propagate~\cite{He2016_CVPR}.
\revs{In particular, the residual connection for a block adds its output with its input and then normalizes this sum as the new output.}

\subsubsection{Position Regressor}
We employ a multi-layer perceptron~(MLP) as the position regressor $\bm \xi$.
This is because MLPs are one of the most widely used and fundamental DNNs for classification and regression tasks and are proven to be universal approximators for arbitrary functions~\cite{Chen1995Universal}.

In particular, the employed MLP comprises two dense encoder blocks. 
It maps $\bm y_n[\tau]\in\mathbb R^{L}$, i.e., each row vector of the output matrix of $\bm \chi$, first to an intermediate vector with half the original length and then to the estimated UE position, i.e., ${\bm p}_{\bm f, n}[\tau]\in \mathbb R^3$.
Finally, combining ${\bm p}_{\bm f, n}[\tau]$ with its corresponding sampling time, i.e., $t_n[\tau]$, the estimated UE track ${\mathcal T}_{\bm f}$ is obtained.

\subsection{Adversarial-learning-based Training Algorithm}
\label{s2ec: train alg track func}

In~(P1), the remaining challenge is to determine the divergence term $\hat{d}_{\tilde{\mathcal Z}}(\mathcal F^{\ssrc}(\bm h), \mathcal F^{\stgt}(\bm h))$.
Accroding to~\eqref{equ: divergence}, it represents the supremum of the empirical probability for an indicator function in $\tilde{\mathcal Z}$ to perform differently on $\mathcal F^{\ssrc}(\bm h)$ and $\mathcal F^{\stgt}(\bm h)$.
To tackle this challenge, based on~\cite{Ganin2016JMLR_Domain,Zhang2019ICML_Bridging}, we adopt a surrogate for the divergence term: the maximal one-minus-cross-entropy~(CE)-loss of a \emph{soft domain classifier} $\zeta:\mathbb R^{L}\rightarrow [0,1]$ in determining the probability for a feature vector to be in $\mathcal F^{\ssrc}(\bm h)$.
This surrogate is intuitive since one-minus-CE-loss is proportional to the extent that the feature vectors of $\mathcal F^{\ssrc}(\bm h)$ and $\mathcal F^{\stgt}(\bm h)$ are distinguishable, which is in accordance with the physical meaning of divergence.

Due to MLPs being universal function approximators~\cite{Chen1995Universal}, we adopt an MLP with the architecture identical to position regressor $\bm \xi$ (except for the output layer) as $\zeta$.
Specifically, given $\bm h$, the CE loss for domain classifier $\zeta$ is defined as:
\beq
\sCE(\zeta,\bm h) = 
- \hspace{-.5em}
\sum\limits_{\bm y \in \mathcal F^{\ssrc}(\bm h)}\!  \frac{\log(\zeta(\bm y))}{|\mathcal F^{\ssrc}(\bm h)|}
- \hspace{-.5em}
\sum\limits_{\bm y \in \mathcal F^{\stgt}(\bm h)}\hspace{-.5em} \frac{\log(1\!-\!\zeta(\bm y))}{|\mathcal F^{\stgt}(\bm h)|}. \nonumber
\eeq

Therefore, denoting the trainable parameters of $\bm h$, $\bm \xi$, $\bm \xi'$, and $\zeta$ by $\bm \theta_{\bm h}$, $\bm \theta_{\bm \xi}$, $\bm \theta_{\bm \xi'}$, and $\bm \theta_{\zeta}$, respectively, $\bm f^*$ in~(P1) can be obtained by solving the following optimization:
\begin{align}
\label{equ: param opt prob}
(\bm \theta_{\bm \xi}^*, \bm \theta_{\bm h}^*) &\! = \! \mathop{\arg}\limits_{(\bm \theta_{\bm \xi}, \bm \theta_{\bm h})} 
\min_{\bm \theta_{\bm \xi}, \bm \theta_{\bm \xi'}, \bm \theta_{\bm h}} \! 
\rev{\Big(}\revs{\hat{\varepsilon}^{\stgt}(\bm \xi \circ \bm h)} \\ 
&+
\revs{\hat{\varepsilon}^{\ssrc}(\bm \xi' \circ \bm h)}  
+
S \max_{\bm \theta_{\zeta}}(1-\revs{\sCE(\zeta,\bm h)}) \rev{\Big)}. \nonumber 
\end{align}

\rev{In~\eqref{equ: param opt prob}, an adversarial maximization exists inside the minimization.
To apply efficient gradient-descent-based DNN optimizers such as Adam~\cite{Kingma2014Adam} for solving~\eqref{equ: param opt prob}, the inner maximization needs to be integrated into the outer minimization.
Nevertheless, directly transforming the inner maximization to the minimization of $\revs{\sCE(\zeta,\bm h)}$ will result in a nonequivalent gradient of the objective function w.r.t. $\bm \theta_{\bm h}$.
To handle the adversarial maximization, we adopt the \emph{gradient reverse function} $\bm R(\cdot)$ in the adversarial learning techniques~\cite{Ganin2016ICML_Unsupervised}.}
Then, the unified minimization can be expressed as
\begin{align}
\label{equ: unify min}
(\bm \theta_{\bm \xi}^*, \bm \theta_{\bm h}^*) &\!=\! \mathop{\arg}\limits_{(\bm \theta_{\bm \xi}, \bm \theta_{\bm h})} 
\min_{\bm \theta_{\bm \xi}, \bm \theta_{\bm \xi'}, \bm \theta_{\bm h}, \bm \theta_{\zeta}} \! 
\revs{\hat{\varepsilon}^{\stgt}(\bm \xi \circ \bm h)}  \\
&+ \revs{\hat{\varepsilon}^{\ssrc}(\bm \xi' \circ \bm h)} 
+ S\cdot \revs{\sCE(\zeta,\bm R(\bm h))}. \nonumber
\end{align}
In~\eqref{equ: unify min}, $\bm R(\cdot)$ satisfies that $
\bm R(\bm h(\mathcal J)) = \bm h(\mathcal J)	$ and $\frac{\partial \bm R(\bm h(\mathcal J))}{\partial \bm \theta_{\bm h}} = -\frac{\partial \bm h(\mathcal J)}{\partial \bm \theta_{\bm h}}$, 
which can be readily implemented by DNN programming.
We refer to~\eqref{equ: unify min} as the \emph{cross-domain training} as it optimizes $\bm f$ for the target domain by using the labeled data for the source domain.
It can be verified that solving both~\eqref{equ: param opt prob} and~\eqref{equ: unify min} with gradient-descent-based DNN optimizers results in equivalent directions for parameter updates.
\rev{Based on~\cite{Lee16COLT_Gradient,Du19ICML_Gradient}, using gradient-descent-based DNN optimizers can lead to local minima of~\eqref{equ: unify min} that are close to its global minimum.}

Furthermore, to speed up the training of $\bm f$ and prevent it from being stuck at poor local minima, we propose a \emph{two-step pre-training} approach.
\revs{As the source domain bears similarities with the target domain, the two-step pre-training finds an efficient initial point for~\eqref{equ: unify min} by solving the tracking error minimization in the source domain.}
In particular, we pre-train the $N_{\sB}$ frame-level encoders in the first step and then pre-train the complete $\bm f$ in the second step.
For Band $n$, the frame-level pre-training can be expressed as:
\begin{align}
\label{equ: pretrain prob}
\mathop{\min}\limits_{\bm \psi_{n}^{\sCSI}, \bm \psi_{n}^{\sDT}, \bm \psi_{n}^{\sRIS},\bm \xi} \hat{\varepsilon}_n^{\ssrc}(\bm \xi \circ \bm \psi_{n}^{\sCSI}\circ( \bm \psi_{n}^{\sDT} \oplus \bm \psi_{n}^{\sRIS} )).
\end{align}
Here, $\hat{\varepsilon}^{\ssrc}_n(\cdot)$ denotes the empirical axial tracking error for the source domain when positioning directly with individual CSI samples.
\rev{Subsequently, in the sequence-level pre-training, we adopt the resulting frame-level encoders of~\eqref{equ: pretrain prob} as the initial point and solve $\min_{\bm f} \hat{\varepsilon}^{\ssrc}(\bm f)$.}
The resulting $\bm f$ is then adopted as the initial point for the cross-domain training of~\eqref{equ: unify min}.

In summary, the complete algorithm for solving~(P1) of the \nname\ framework is presented as Algorithm~\ref{alg: x2trans framework}.

\textbf{Complexity Analysis}: 
We derive the computational complexity of training $\bm f$ in Algorithm~\ref{alg: x2trans framework} w.r.t. the important hyperparameters, including $Q_n$, $ N_{\sSC,n}$, $M_n$, and $F_n$, $\forall n\in\{1,\cpdots,\! N_{\sB}\}$, in order to evaluate its scalability for larger systems.
For Band $n$, the complexity of the frame-level encoder is $\mathcal O((Q_n N_{\sSC,n}M_n)^2)$, dominated by that of the first dense encoder block.
Then, as $|\mathcal I_n|$ is proportional to $F_n$, encoding $\mathcal I_n$ at the frame-level is of complexity $\mathcal O(F_n(Q_n N_{\sSC,n}M_n)^2)$.
For $\bm \chi$, its complexity is $\mathcal O((\sum_n F_n)^2)$, dominated by that of the MHSA blocks.
For $\bm \xi$ and $\zeta$, their computational complexity is $\mathcal O(\sum_n F_n)$.
Therefore, the complexity of $\mathcal T_{\bm f} = \bm f(\mathcal J)$ is $\mathcal O((\sum_n \! F_n)^2\!+\!\sum_{n} \!F_n(Q_n N_{\sSC,n}M_n)^2)$. 
As for the gradient calculation of the trainable parameters of $\bm f$, the amount of computation scales linearly with that of $\mathcal T_{\bm f} = \bm f(\mathcal J)$~\cite[Ch. 6.5]{goodfellow2016deep}.
Consequently, the complexity of training $\bm f$ is $\mathcal O((\sum_n \! F_n)^2\!+\!\sum_{n} \!F_n(Q_n N_{\sSC,n}M_n)^2)$, determined by the squared total number of frames in a tracking period and the squared dimension of the RIS-CSI.

\rev{\textbf{Generalizability Analysis}: The designed algorithm under the \name\ framework possesses the following generalizability: \emph{First}, instead of requiring any theoretical relationship established between CSI and UE track, \name\ enables the algorithm to learn their relationship automatically, so that the algorithm generalizes to more intricate and complex wireless channel conditions.
\emph{Second}, as the hierarchical architecture in \name\ decomposes the signal processing of each CSI-modality in each frequency band through the modal-specific CSI encoders and frame-level encoders, the algorithm readily generalizes to systems with more RISs and more frequency bands by employing additional encoders.
\emph{Third}, thanks to the cross-domain learning of \name, the algorithm is able to generalize its positioning capability trained in a source environment to different target deployment environments, using only sparse labeled CSI data.}

\begin{figure}[!t]
\vspace{-0.7em}
\begin{algorithm}[H]
\small
\caption{Cross-domain training with two-step pre-training for solving (P1) of \name.}
\label{alg: x2trans framework}
\begin{algorithmic} [1]
\State Initialize $\bm f$ as designed in Sec.~\ref{s2ec: MM CB Transformer}.
\For{$\text{epoch}=1,...,E_{\spre,1}$}\emph{\quad \# Frame-level pre-training}
\For{next $B_{\spre,1}$ labeled data in $\mathcal D^{\ssrc}$}
\State For each Band $n$ ($n\in\{1,\cpdots,\!N_{\sB}\}$), evaluate the objective function of \eqref{equ: pretrain prob} based on the labeled data.
\State For each Band $n$, calculate gradients and update the parameters of $\bm \psi_n^{\sCSI}$, $\bm \psi_n^{\sDT}$, $\bm \psi_n^{\sRIS}$, and $\bm \xi$ with learning rate $\eta$.
\EndFor
\EndFor
\For{$\text{epoch}=1,...,E_{\spre,2}$}\emph{\quad \# Sequence-level pre-training}
\For{next $B$ labeled data in $\mathcal D^{\ssrc}$}
	\State Based on the labeled data, evaluate $\hat{\varepsilon}^{\ssrc}(\bm f)$.
	\State Calculate gradients and update the parameters of $\bm f$ with learning rate $\eta$.
\EndFor
\EndFor
\For{$\text{epoch} = 1,...,E$}\emph{\quad \# Cross-domain training}
\For{next $B$ labeled and unlabeled data in $\mathcal D^{\ssrc}$, $\bar{\mathcal D}^{\ssrc}$, and $\bar{\mathcal D}^{\stgt}$}
	\State Randomly draw $B$ labeled data from $\mathcal D^{\stgt}$.
	\State Based on the labeled and unlabeled data, evaluate the objective function of \eqref{equ: unify min}.
	\State Calculate gradients and update $\bm \theta_{\bm \xi}, \bm \theta_{\bm \xi'}, \bm \theta_{\bm h}, \bm \theta_{\zeta}$ with learning rate $\eta$.
\EndFor
\EndFor
\end{algorithmic}
\end{algorithm}
\vspace{-1.5em}
\end{figure}

\section{Simulation Results}
\label{sec: evaluation}

\subsection{Simulation Setup}\label{s2ec: simul set}

We consider an $(x,y,z)$-coordinate system with the origin at the BS's base.
In this coordinate system, ROI is defined as a 3D cuboid region $\mathcal A = \{(x,y,z)~\!\text{m}| x\in[0,100], y\in[- 50,50], z\in[0,10]\}$.
To facilitate presentation, we focus on the case where $N_{\sB}=2$.
\revs{Based on the 3GPP standard for 5G NR~\cite{3GPP_TS_38_101}, we select a sub-6GHz band and an mmWave band, centered at~$5.9$~GHz and $28.0$~GHz, with bandwidths of $20$~MHz and $120$~MHz and $N_{\sSC,1}=24$ and $N_{\sSC,2}=66$ sub-carriers, respectively.}
For Bands $1$ and $2$, the centers of Rx antennas are at $(0, 0, 22)$~\!m and $(0, 0, 20)$~\!m, respectively.
The Rx antennas are spaced at half-wavelength intervals of the center frequencies and arranged along the $y$-axis.
The centers of the RISs for Bands $1$ and Band $2$ are located at $(50, 50, 22)$~\!m and $(50, 50, 20)$~\!m, respectively, with their normal directions pointing in the opposite direction of the $y$-axis.
To reduce the storage and computation complexity, we sample one out of every two Rx antennas and one out of every $3\times 3$ meta-elements when generating the CSI indicators.

In a tracking period, each UE starts at a random position within $\mathcal A$ and moves at a speed evenly sampled from $[0,10]$~\!m/s in a random direction.
For Bands~$1$ and~$2$, the centers of the Tx antennas are positioned 7~\!cm below and above the UE's center, respectively.
Besides, the probabilities of a scheduled UL transmission in Bands $1$ and $2$ are $P^{\sUL}_1=P^{\sUL}_2=0.1$.
In addition, we adopt the channel model in the 3GPP standard for urban macro deployment scenarios~\cite{3GPP_TR_38_901}:
For the channel from a UE to the Rx antennas (or an RIS), the probability of an LoS path is $\mathrm{Pr}=\min(\frac{18}{d_{\serD}} + \exp(-\frac{d_{\serD}}{63})(1-\frac{18}{d_{\serD}}),1)$ with $d_{\serD}$ being the horizontal distance between the Tx and the Rx antennas~(or the RIS's center);
and the variance of the aggregate gain of environmental scattering paths for DT (or RIS) channel is $\varLambda_n = 10^{-3.24}(f_{n}/10^{9})^{-2}d^{-3}$ with $f_n$ being the center frequency of Band $n$ and $d$ being the distance between the Tx and the Rx antennas~(or the RIS's center).
The reflection coefficient variances are $V_n^{\sDT}\!=\!V_n^{\sRIS}\!=\!0.1$, resulting in strong environmental reflection interference.

As for the measurement noise in~\eqref{equ: csi format with noise}, we by default set \rev{variances $\sigma_n^{\sDT}$ and $\sigma_n^{\sRIS}$}
for Bands $1$ and $2$ to be the same, satisfying that the SNR of RIS-CSI for Band $2$ at the center of  $\mathcal A$ is $\mathrm{SNR}^{\sRIS}_{\scen, 2}=10$~\!dB.
Furthermore, the generated datasets are divided into training and test sets at a ratio of $9:1$. 
By default, the axial tracking errors presented in the following evaluation results are for the test sets.
The other default parameters are summarized in Table~\ref{table: parameter}.

\begin{table}
\centering
\caption{Simulation Parameters}
\label{table: parameter}
\setlength{\extrarowheight}{3pt}
\begin{scriptsize}
\begin{tabular}{||p{1.4cm}|p{.8cm}||p{1cm}|p{.5cm}||p{1.4cm}|l||}
\hline
\textbf{Parameter}     & \textbf{Value}& 
\textbf{Parameter}     & \textbf{Value}&
\textbf{Parameter}     & \textbf{Value}
\\ \hline
$T_{\sTP}$ & $500$~\!ms &
$T_{n}$ & $10$~\!ms &
$F_n$ & $50$ 
\\ \hline
$M_n$ & $12\!\times\! 12$ &
$N_{\sRx, n}$ & $4$ &
$N_{\sTx, n}$ & $2$  
\\ \hline
$\gamma$ & $0.3$ & 
$R,R'$ & $10$ &
$|\mathcal D^{\ssrc}|$,$|\bar{\mathcal D}^{\ssrc}|$ & $10^4$
\\ \hline
$|\mathcal D^{\stgt}|$ & $500$ &
$|\bar{\mathcal D}^{\stgt}|$ & $10^4$ &
$L$ & $128$
\\ \hline
$L_{\sTM}$ & $32$ & 
$C$ & $8$  & 
$A$ & $16$ 
\\ \hline

$\eta$ & $0.001$ &
$B_{\spre,1}$ & $256$ & 
$B$ & $32$ 
\\ \hline
$E$ & $50$ &
$E_{\spre,1}$ & $10$ & 
$E_{\spre,2}$ & $200$ 
\\ \hline
\end{tabular}
\end{scriptsize}
\vspace{-.5em}
\end{table}

\subsection{Evaluation Results}
\label{ssec:eval_result}

\emph{Firstly}, we verify \nname's general capability of tracking UE with CSI indicators.
To facilitate presentation and comparison, we first consider the case where the target and source domains are the same and evaluate the training performance of $\bm f$ in terms of the average axial tracking errors for training and test sets.
\begin{figure}[t]  
  \centering
  \includegraphics[width=.65\linewidth]{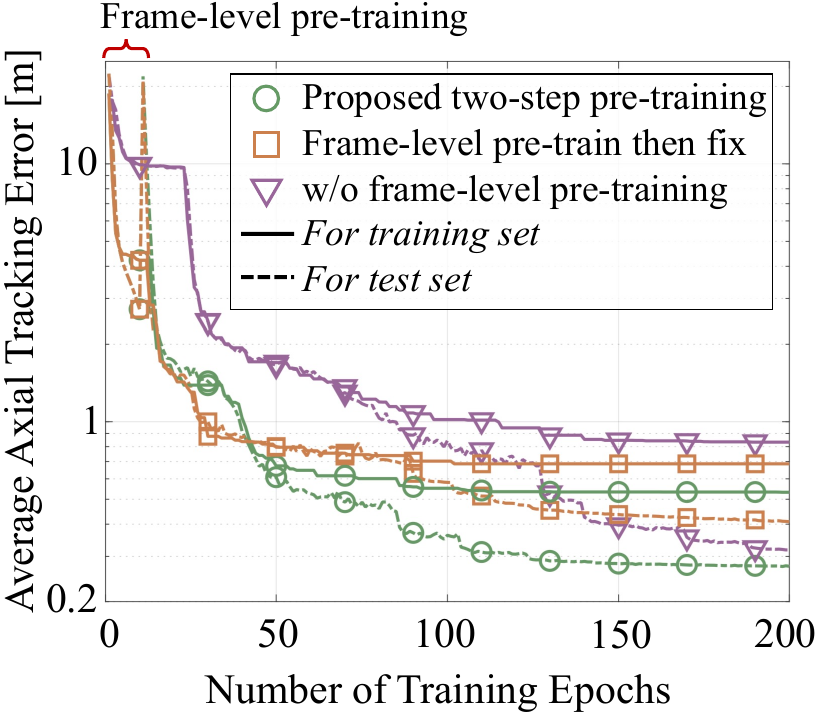}
    \vspace{-.5em}
    \caption{\revs{Average axial tracking errors of $\bm f$ for training and test sets versus the number of training epochs, given different training algorithms.}}
    \label{fig: compare pretrain}
  \vspace{-1.2em}
  \end{figure}
As shown in Fig.~\ref{fig: compare pretrain}, when the number of training epochs increases, the average axial tracking errors for training and test sets decrease and converge to the decimeter level, proving the effectiveness of the proposed hierarchical architecture and its implementation in Secs.~\ref{s3ec: model of f} and~\ref{s2ec: MM CB Transformer}, respectively. 
Besides, the pre-training of the frame-level encoders in Algorithm~\ref{alg: x2trans framework} enables the training of $\bm f$ to converge significantly faster and to a lower average axial tracking error.
Even if the frame-level encoders are fixed after pre-training, a low average axial tracking error can still be achieved.
Therefore, the training of frame- and sequence-level encoders can be potentially separated to avoid the high complexity of training the complete $\bm f$.
This is helpful when more sophisticated DNN blocks are employed to handle CSI indicators with larger dimensions and/or for more bands.

\begin{figure}[b]  
  \vspace{-1.2em}
\centering
\includegraphics[width=1\linewidth]{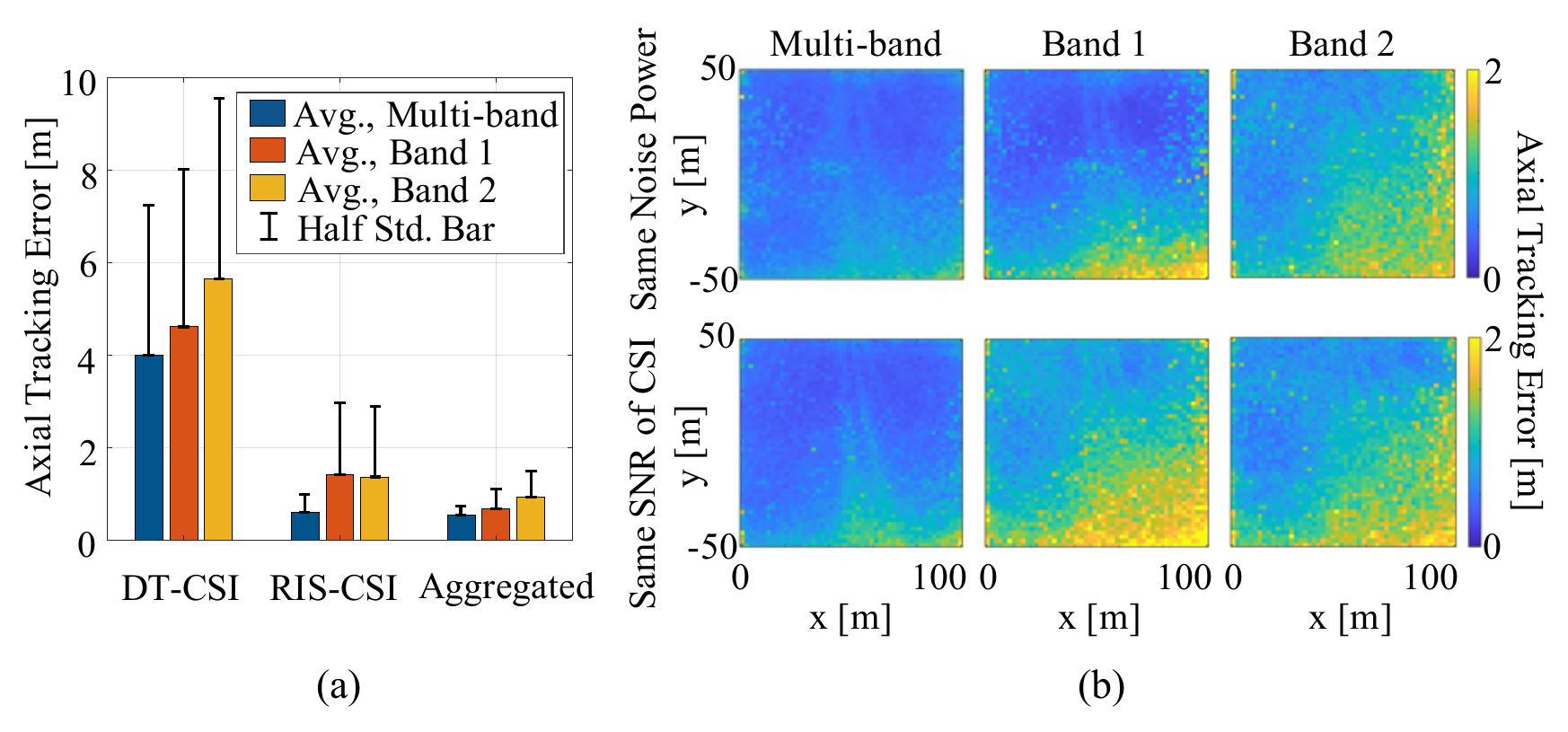}
  \vspace{-1.6em}
  \caption{(a) Avg. and Std. of axial tracking errors for using CSI indicators of different modalities and different bands; (b) Horizontal distributions of axial tracking errors for using CSI indicators of different bands under the same measurement noise power and the same SNR of CSI indicator conditions.}
  \label{fig: compare CSI modal}
\end{figure}

\emph{Secondly}, we evaluate \nname's capability of utilizing multi-modal CSI and  handling the irregularity and asynchrony of MB CSI sequences.
In Fig.~\ref{fig: compare CSI modal}~(a), we compare the cases where either DT-CSI or RIS-CSI is used solely and where DT- and RIS-CSI are aggregated as in \nname.
\rev{Intuitively, the DT-CSI case represents the scenario without RISs, and the RIS-CSI case represents the scenario where only CSI indicators of RISs are available due to the blockage of direct transmission paths.}
It can be observed that \nname\ efficiently combines the DT-CSI and RIS-CSI and achieves lower average~(Avg.) and standard deviation~(Std.) of axial tracking errors: $11\%$ lower Avg. and $49\%$ lower Std. than those for only RIS-CSI.
Besides, we compare the cases where CSI sequences for Band~1 or Band~2 are used solely and where they are used jointly as MB CSI sequences.
It can be observed that \nname\ jointly utilizes the CSI sequences across the two bands and obtains higher accuracy and precision in UE tracking: $20\%$ lower Avg. and $54\%$ lower Std. than those for only Band~$1$.

Then, in Fig.~\ref{fig: compare CSI modal}~(b), we compare the spatial distribution of axial tracking errors in the horizontal plane.
By using MB CSI sequences, \nname\ reduces axial tracking errors especially in areas far from the BS and RISs.
To derive more insights, we compare the axial tracking errors of $\bm f$ under different conditions of the measurement noise for CSI indicators.
The first row of Fig.~\ref{fig: compare CSI modal}~(b) is obtained in the default case where the measurement noise power for the two bands is the same, satisfying $\mathrm{SNR}^{\sRIS}_{\scen, 2}\!=\!10$~\!dB.
In this case, the axial tracking errors for Band~$1$ are smaller as a lower center frequency leads to lower attenuation and a higher SNR for CSI indicators.
The second row is obtained when the RIS-CSI measurement at the center of $\mathcal A$ are of the same SNR for the two bands: $\mathrm{SNR}^{\sRIS}_{\scen, 1}\!=\!\mathrm{SNR}^{\sRIS}_{\scen, 2}\!=\!5$~\!dB.
In this case, the axial tracking errors for Band~$2$ are lower as CSI indicators of a larger number of subcarriers contain richer positional information.

\begin{figure}[t] 
\centering
\includegraphics[height=9.4em]{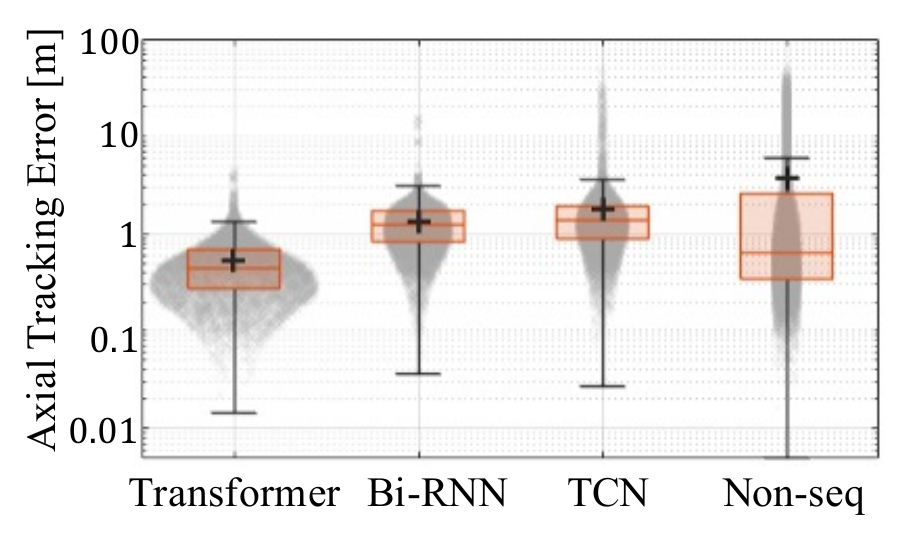}
  \vspace{-.6em}
    \caption{Distributions of the axial tracking errors for the tracking functions employing different sequence-level encoders. ``+'' indicates the mean value.}
    \label{fig: compare seq modal}
 \vspace{-1.3em}
\end{figure}
\emph{Thirdly}, we conduct benchmark comparisons for Algorithm~\ref{alg: x2trans framework}.
To the best of the authors' knowledge, there are no existing comparable algorithms that can handle multi-modal CSI indicators and achieve mutual complementation among irregular and asynchronous MB CSI sequences.
Therefore, we focus on justifying our transformer-based $\bm f$ by comparing it with those employing other benchmark DNNs, including bi-directional recurrent neural network~(Bi-RNN) based on gated recurrent units~(GRU)~\cite{Chung2014NIPS_Empirical} and temporal convolutional network~(TCN) based on dilated CNN blocks~\cite{bai2018empirical}.
For a fair comparison, the Bi-RNN and TCN are of the same number of cascaded blocks as the transformer: The Bi-RNN is composed of $8$ pairs of forward and backward GRU layers, and the TCN is composed of $8$ double-layered dilated 1D CNN blocks with residual connections.
As shown in Fig.~\ref{fig: compare seq modal}, the proposed transformer-based $\bm f$ outperforms the others, which can be explained as follows:
Compared to the Bi-RNN that handles input feature vectors iteratively, the MHSA blocks of transformer handle them parallelly.
This makes the transformer less susceptible to the vanishing gradient problem, especially for long input sequences.
Besides, the transformer is more efficient in handling MB CSI sequences than the TCN because the dilated CNN blocks in the TCN rely on regularly-spaced and synchronous input sequences to recognize temporal patterns.

Moreover, we show the axial tracking errors without sequence-level encoders in the \emph{Non-seq} case in Fig.~\ref{fig: compare seq modal}.
It can be observed that processing the CSI feature vectors with the sequence-level encoder can significantly reduce the average and variance of axial tracking errors.
This is because the sequence-level encoder can achieve mutual complementation, handling the absence of LoS paths and reducing the strong interference of environmental reflection paths.

\begin{figure}[t] 
\centering
\includegraphics[width=1\linewidth]{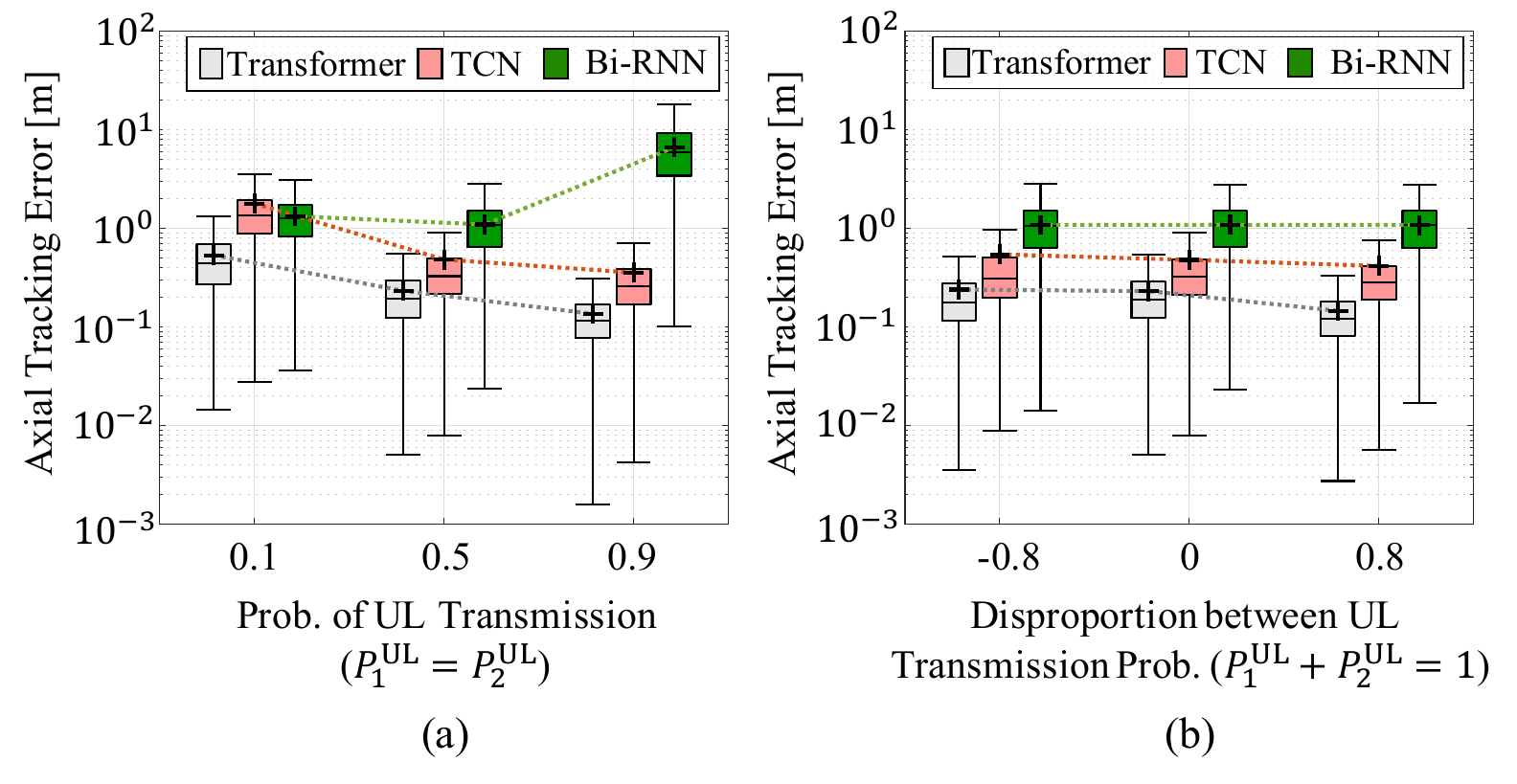}
  \vspace{-2em}
    \caption{Box plots of the axial tracking errors for the tracking functions with different sequence-level encoders, under different UL traffic conditions for the two bands. (a)~Same UL transmission probabilities; (b)~Disproportional UL transmission probabilities. ``+'' indicates the mean value.}
    \label{fig: compare sparsity}
\vspace{-1.2em}
\end{figure}

\emph{Fourthly}, we compare the axial tracking errors under different conditions of UE's UL traffic.
Fig.~\ref{fig: compare sparsity}~(a) shows the axial tracking errors versus the probabilities~(Prob.) of UL transmission in Bands $1$ and $2$, i.e., $P^{\sUL}_1$ and $P^{\sUL}_2$, given $P^{\sUL}_1=P^{\sUL}_2$.
As $P^{\sUL}_n$~($n=1,2$) increases, the axial tracking errors for transformer-based and TCN-based $\bm f$ decrease because the increased number of CSI samples provide more positional information of the UE.
Even with scarce UL data traffic, i.e., $P^{\sUL}_n=0.1$, the proposed $\bm f$ can achieve an average axial tracking error of around $0.5$~\!m, which can be further reduced to $0.11$~\!m when $P^{\sUL}_n=0.9$.
However, the average axial tracking error for the Bi-RNN-based $\bm f$ rises when $P^{\sUL}_n$ increases from $0.5$ to $0.9$.
This is because longer MB CSI sequences exacerbate the vanishing gradient problem.
Besides, Fig.~\ref{fig: compare sparsity}~(b) shows the axial tracking errors versus the disproportion between UE's UL transmission probabilities in Bands~$1$ and~$2$, i.e., $P^{\sUL}_1-P^{\sUL}_2$.
It can be observed that the performance of all the tracking functions is generally robust to disproportional UL traffic.
The decreasing trends for the axial tracking error w.r.t. $P^{\sUL}_1-P^{\sUL}_2$ is because the CSI indicators for Band $1$ have higher SNR in the default setting, as shown in Fig.~\ref{fig: compare CSI modal}~(b), leading to lower axial tracking errors.

\emph{Fifthly}, we verify the \nname's cross-domain learning capability when the target domain is different from the source domain.
\rev{As demonstrated in Fig.~\ref{fig: compare CSI modal}, the tracking precision is mainly contributed by the RIS-CSI.
Thus, we focus on the change in the RISs' positions for the target domain, which is expected to have a substantial impact on the tracking errors, facilitating the evaluation of the cross-domain learning capability of \name.}
Specifically, the source domain corresponds to the environment with the default setting, and the target domain corresponds to the environment where the two RISs are moved to $(50, -50, 22)$~\!m and $(50,-50, 20)$~\!m, which are distinctly different from their default sites.
We compare \nname\ with other benchmark approaches for DA, including fine-tuning, unsupervised DA~(UDA), supervised DA~(SDA), and semi-supervised DA~(Semi-SDA):
\begin{figure}[t] 
    \centering
    \includegraphics[height=12.4em]{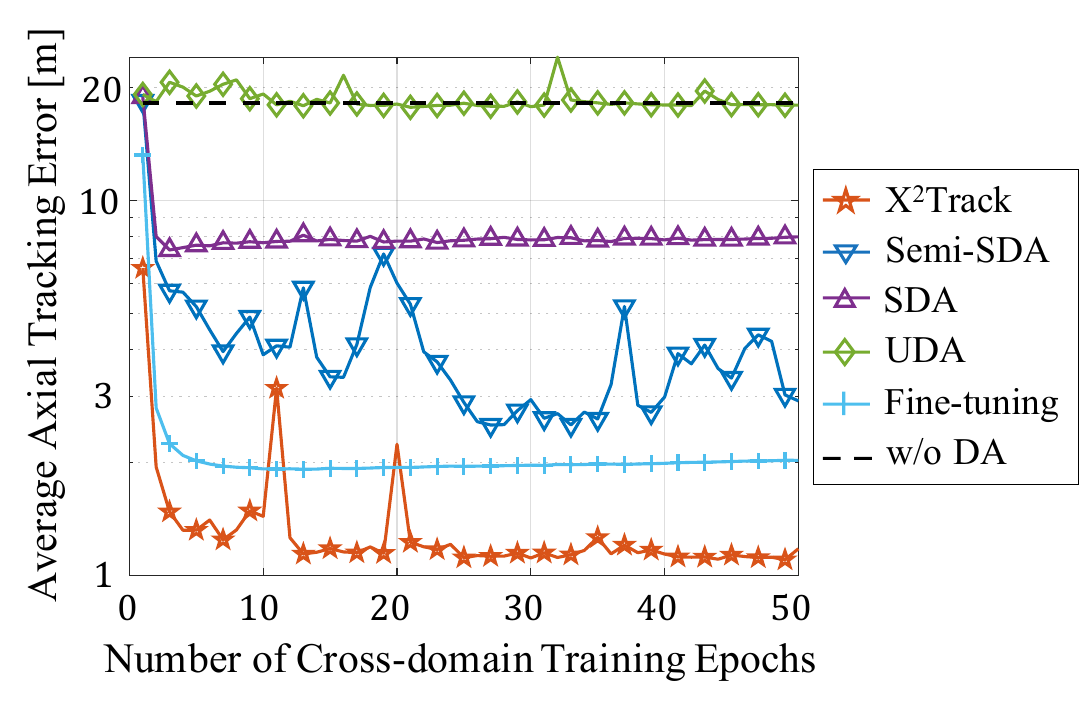}
        \caption{Comparison between \nname\ and other benchmark approaches for DA in terms of the average axial tracking errors versus the number of cross-domain training epochs after the pre-training for the source domain.}
        \label{fig: domain adaptation methods}
    \vspace{-.9em}
    \end{figure}

\begin{enumerate}
\item \textbf{Fine-tuning}~\cite{goodfellow2016deep}: After the two-step pre-training for the source domain, minimize $\hat{\varepsilon}^{\stgt}(\bm f)$ w.r.t. $\bm f$ with the learning rate for $\bm h$ being reduced to $5\times 10^{-5}$.
\item \textbf{UDA}~\cite{Ganin2016ICML_Unsupervised, Zhang2019ICML_Bridging}: Minimize $\hat{\varepsilon}^{\ssrc}(\bm \xi\circ\bm h)+S\cdot\sCE(\zeta, \bm R(\bm h))$ w.r.t. $\bm \xi$, $\bm h$, and $\zeta$ without utilizing $\mathcal D^{\stgt}$.
\item \textbf{SDA}~\cite{Ben_David2010ATheory}: Minimize $\hat{\varepsilon}^{\ssrc}(\bm f)+\hat{\varepsilon}^{\stgt}(\bm f)$ w.r.t. $\bm f$ without utilizing $\bar{\mathcal D}^{\stgt}$.
\item \textbf{Semi-SDA}~\cite{Li2021CVPR_Learning}: Minimize $\hat{\varepsilon}^{\ssrc}(\bm \xi\circ\bm h)+\hat{\varepsilon}^{\stgt}(\bm \xi\circ\bm h)+S\!\cdot\!\sCE(\zeta, \bm R(\bm h))$ w.r.t. $\bm \xi$, $\bm h$, and $\zeta$.
\end{enumerate}

Fig.~\ref{fig: domain adaptation methods} shows that the cross-domain learning of \nname\ by the conversion of $\varepsilon^{\stgt}(\bm f)$ leads to the lowest average axial tracking error for the target domain, compared with the benchmarks.
In Fig.~\ref{fig: domain adaptation methods}, UDA cannot effectively adapt the tracking function to the target domain because, even though $\bm h$ is trained to get indistinguishable position feature vectors for the source and target domains, it lacks effective adaptation of $\bm \xi$. 
In this regard, SDA achieves better cross-domain learning than UDA by involving the minimization of $\hat{\varepsilon}^{\stgt}(\bm f)$.
Compared to the SDA, Semi-SDA leads to a lower average axis tracking error because it utilizes the large amount of unlabeled data in $\bar{\mathcal D}^{\stgt}$ by minimizing the divergence surrogate, reducing the overfitting to the small labeled dataset $\mathcal D^{\stgt}$.
On the other hand, fine-tuning adapts $\bm \xi$ to the target domain without considering its performance on $\mathcal D^{\ssrc}$ and thus reduces the axial tracking errors to a larger extent compared to the others.
This supports the separation of position regressors for the source and target domains.
By jointly minimizing empirical axial tracking errors for the source and target domains and the divergence surrogate with separated position regressors, \nname\ outperforms all the benchmarks, reducing the average axial tracking error by $42\%$ compared to the second-best.

\emph{Finally}, we evaluate the performance of \nname\ for various target deployment environments.
\rev{In practice, typical differences between target deployment environments and the source environment exist in three aspects: i) the \emph{characteristics of wireless channels}, ii) the \emph{positions of BSs and RISs}, and iii) the \emph{characteristics of UEs}.}
\rev{For the first aspect, we consider three representative cases of target environments with increased levels of multipath interference and noise, which have stronger reflection paths~(SRP), more reflection paths~(MRP), and higher measurement noise~(HMN), respectively.}
\rev{For the second aspect, we consider two cases, which have higher BSs~(HBS) and nearer RISs' locations~(NRL), respectively.
For the third aspect, we consider three cases, which have bigger UEs~(BUE), faster UEs~(FUE), and chaotic UE tracks~(CUT), respectively.
The detailed descriptions and specifications of the above eight cases are listed in Table~\ref{table: tgt envir}.}

\begin{figure}[t] 
\centering
\includegraphics[width=1\linewidth]{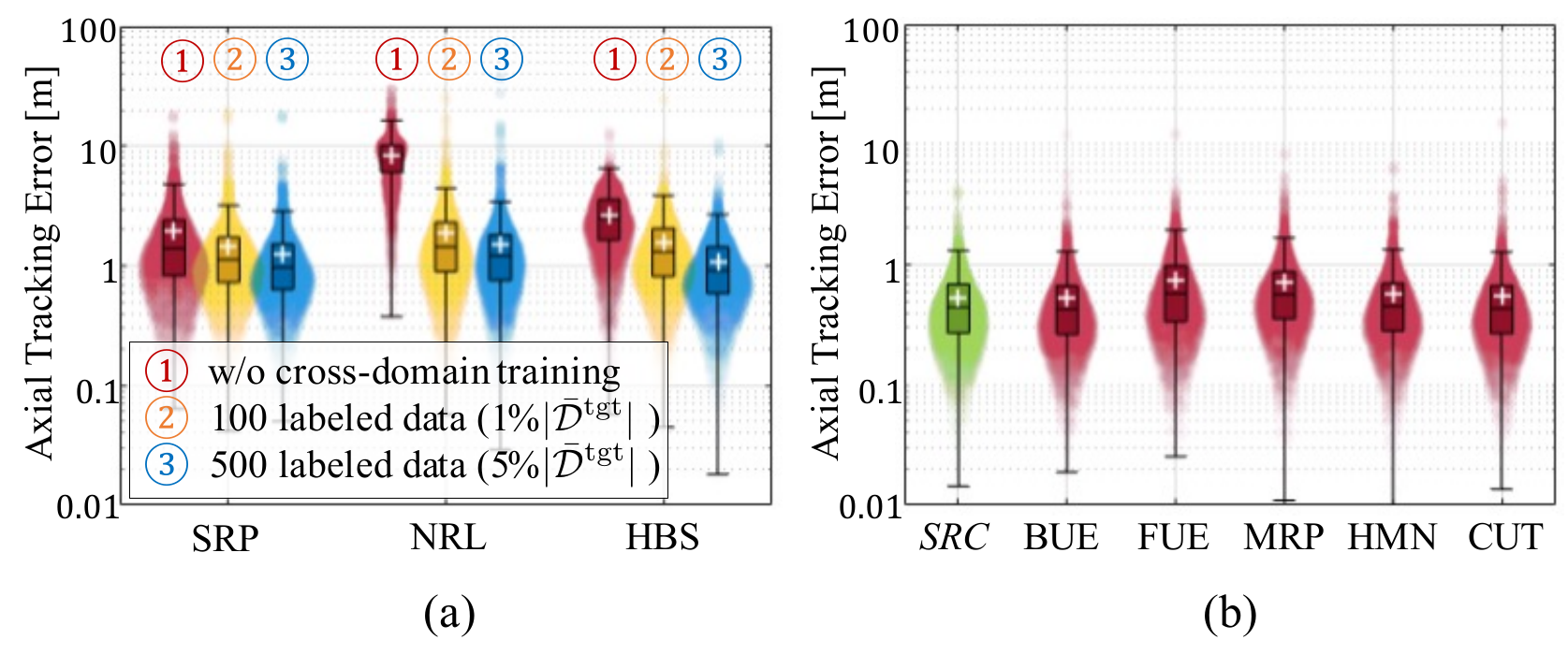}
  \vspace{-1.5em}
    \caption{Box plots of axial tracking errors (a)~for the SRP, NRL, and HBS cases, where the cross-domain training in Algorithm~\ref{alg: x2trans framework} is used; and (b)~for source domain~(\emph{SRC}) and BUE, FUE, MRP, HMN, and CUT cases, where the tracking function after the two-step pre-training in Algorithm~\ref{alg: x2trans framework} is used.}
    \label{fig: various target domain}
\vspace{-.5em}
\end{figure}

It can be observed in Fig.~\ref{fig: various target domain}~(a) that, for SRP, NRL, and HBS cases, with sparse labeled data~($|\mathcal D^{\stgt}|=5\%|\bar{\mathcal D}^{\stgt}|=500$), \nname\ achieves an average axis tracking error close to $1$~m in the target deployment environments, reducing $35\%\!\sim\!82\%$ axial tracking error increment due to the difference between source and target domains.
Furthermore, under the \nname\ framework, the pre-trained $\bm f$ for $\mathcal D^{\ssrc}$ shows inherent robustness towards the difference between source and target domains in BUE, FUE, MRP, HMN, and CUT cases, requiring no additional efforts for cross-domain learning.
In Fig.~\ref{fig: various target domain}~(b), it can be observed that the tracking function trained with $\mathcal D^{\ssrc}$ achieves low axial tracking errors for these target domains with an average error increment less than~$0.22$~\!m.

\begin{table}
\centering
\caption{Representative Target Deployment Environments}
\label{table: tgt envir}
\setlength{\extrarowheight}{2pt}
\definecolor{lightgray}{gray}{0.9}
\begin{scriptsize}
    \begin{tabular}{||m{0.2\textwidth}|m{0.23\textwidth}||}
    \hhline{|-|-|}
    \cellcolor{lightgray}\textbf{Name}     &{\cellcolor{lightgray}\textbf{Difference with the Default Source Environment}}
    \\ \hline
    Stronger Reflection Paths~(SRP) & \rev{Environmental reflectors and scatterers have larger reflection coefficients}: the variances of reflection coefficients, i.e., $V_n^{\sDT}$ and $V_n^{\sRIS}$, are $5$ times larger.
    \\ \hline
    Nearer RISs' Locations~(NRL) & \rev{The RISs are closer to the BSs}: the $x$-coordinates of the RISs are reduced by $15$~\!m.
    \\ \hline
    Higher BSs~(HBS) & \rev{The heights of the BSs' positions are increased}: the $z$-coordinates of the Rx antennas are increased by $10$~\!m.
    \\ \hline
    Bigger UEs~(BUE) & \rev{The model of UEs has a bigger size}: the relative distance from a UE's Tx antennas to the center of the UE is doubled.
    \\ \hline
    Faster UEs~(FUE) & \rev{In the target environment, the UEs move at faster speeds}: the speed range of UEs is doubled.
    \\ \hline
    More Reflection Paths~(MRP) & \rev{The target environment has more reflectors}: the numbers of environmental reflection paths, i.e., $R$ and $R'$, are doubled.
    \\ \hline 
    Higher Measurement Noise~(HMN) & \rev{Measuring CSI suffers from higher noise}: the power of measurement noise for CSI indicators is increased by $5$~\!dB.
    \\ \hline
    Chaotic UE Tracks~(CUT) & \rev{The tracks of UEs are more irregular and chaotic}: the UEs move randomly forwards and backwards within a tracking period.
    \\ \hline
    \end{tabular}
\end{scriptsize}
\end{table}

\section{Conclusion}
\label{sec: conclu}

This paper has investigated the CSI-based UE tracking for RIS-aided MB ISAC systems, where we have formulated the tracking error minimization problem and proposed a novel framework named \nname\ for solving it.
By designing the architecture of the tracking function and converting the intractable objective function, \nname\ explicitly handles the challenges including the diverse multi-modality of CSI indicators, the irregularity and asynchrony of MB CSI sequences, and the sparsity of labeled data for the target deployment environment.
To implement the \nname\ framework efficiently, we have designed a deep learning algorithm based on transformer and adversarial learning techniques.

Simulation results have revealed the following insights:
Firstly, \nname\ efficiently utilizes multi-modal MB CSI sequences, reducing $11\%\!\sim\!20\%$ Avg. and $49\%\!\sim\!54\%$ Std. of axial tracking errors compared to those of single-modal and single-band counterparts.
Secondly, even with scarce UL transmissions and strong environmental reflection interference, \nname\ achieves decimeter-level tracking precision. 
Finally, \nname\ effectively optimizes the tracking function for various target deployment environments, using only sparse labeled data ($500$ labeled tracking periods of $0.5$ seconds, i.e., less than $5$~minutes of labeled UE tracks), and is robust to changes in channel conditions and UE's type and~speed.

\rev{Based on \name, we expect future research can be conducted in the following directions:
\emph{First}, leveraging the capability of tracking a UE, the \nname\ can be extended to support user activity recognition for RIS-aided MB ISAC systems, which can help a BS understand the semantic context of its users to predict their traffic demands and rate requirements.
As a \emph{second} step, recognizing the semantic context of a user can reciprocally enhance the tracking precision for the user, which calls for further research efforts especially in protocol design and performance optimization.}
 

\vspace{-.8em}
\renewcommand{\refname}{References} 

\bibliographystyle{IEEEtran}
\bibliography{bibilio}

\balance
\end{document}